\documentclass{article}
\pdfoutput=1
\usepackage{authblk}
\usepackage[numbers,sort&compress]{natbib}
\usepackage{textcomp}
\usepackage{amsmath}
\usepackage{subfigure}
\usepackage{amssymb}
\usepackage[figuresright]{rotating}
\usepackage{color}
\usepackage{caption}
\usepackage{setspace}
\usepackage[utf8]{inputenc} 
\usepackage[T1]{fontenc}    
\usepackage{hyperref}       
\usepackage{url}            
\usepackage{booktabs}       
\usepackage{amsfonts}       
\usepackage{nicefrac}       
\usepackage{microtype}      
\usepackage{xcolor}         
\usepackage{graphicx}
\usepackage{graphbox}       
\usepackage{pict2e}
\usepackage{xcolor}
\usepackage{wrapfig}
\usepackage{amssymb,amsmath,amsthm}
\usepackage{bm}
\usepackage{diagbox}
\usepackage{mathtools}
\usepackage{algorithm}  
\usepackage[noend]{algpseudocode}
\usepackage{enumerate}
\usepackage[a4paper,left=3cm,right=3cm,top=3cm,bottom=3cm]{geometry}

\theoremstyle{plain}

\theoremstyle{plain}

\theoremstyle{plain}

\theoremstyle{plain}

\providecommand{\assumptionname}{Assumption}
\providecommand{\lemmaname}{Lemma}
\providecommand{\propositionname}{Proposition}
\providecommand{\theoremname}{Theorem}

\title{Semi-supervised Impedance Inversion by Bayesian\\ Neural Network Based on 2-\textit{d} CNN Pre-training}

\author[a]{Muyang Ge}
\author[a]{Wenlong Wang \thanks{Corresponding author: wenlong.wang@hit.edu.cn}}
\author[b]{Wangxiangming Zheng}

\affil[a]{Department of Mathematics, Harbin Institute of Technology}
\affil[b]{Yau Mathematical Sciences Center, Tsinghua University}

\date{}

\begin{document}
\maketitle
\begin{abstract} 
  Seismic impedance inversion can be performed with a semi-supervised learning algorithm, which  only needs a few logs as labels and is less likely to get overfitted. However, classical semi-supervised learning algorithm usually leads to artifacts on the predicted impedance image. In this artical, we improve the semi-supervised learning from two aspects. First, by replacing 1-$d$ convolutional neural network (CNN) layers in deep learning structure with 2-$d$ CNN layers and 2-$d$ maxpooling layers, the prediction accuracy is improved. Second, prediction uncertainty can also be estimated by embedding the network into a Bayesian inference framework. Local reparameterization trick is used during forward propagation of the network to reduce sampling cost. Tests with Marmousi2 model and SEAM model validate the feasibility of the proposed strategy.
\end{abstract}

\section{Introduction}
Traditional strategies for seismic impedance inversion includes full waveform inversion, AVO inversion and functional optimization \cite{tarantola1986strategy,buland2003bayesian, li2017seismic}. Meanwhile, seismic impedance inversion can be seen as a kind of supervised learning problem in machine learning, and a lot research has applied various machine learning algorithm in seismic impedance inversion, including linear and nonlinear regression \cite{russell2019machine} and symbolic regression \cite{nunes2019fast}.

Deep learning is one of the machine learning strategies, which is based on 
deep neural network (DNN). Deep learning has also been widely used in seismic impedance inversion. The use of fully-connected linear layers is proved to be effective in improving the accuracy of inversion result \cite{kim2018geophysical, iturraran2021machine}. Das et al. and Wu et al. utilize convolutional neural network (CNN) to extract features along the seismic traces \cite{das2019convolutional, wu2021deep}. Alfarraj et al. apply recurrent neural network (RNN) in seismic impedance inversion by dealing with seismic traces as time series \cite{alfarraj2018petrophysical}. There are many studies applying other practical deep learning structures in impedance inversion, such as convolutional residual network and transfer learning \cite{wu2020seismic}, generative adversarial network (GAN) \cite{meng2021seismic} and joint learning \cite{mustafa2020joint}.

One tricky problem is that supervised learning needs abundant labeled data for training. To alleviate this problem, Alfarraj and AlRejib propose semi-supervised learning for seismic impedance inversion to infer acoustic impedance (AI) and elastic impedance (EI) \cite{alfarraj2019semi, alfarraj2019semisupervised}. It only needs a few logs as labels and is less likely to get overfitted. A lot subsequent research aims to develop this strategy. Mustafa et al. improve this work by extracting both spatial and temporal features of seismic data profile \cite{mustafa2020spatiotemporal}. Wu et al. and Meng et al. use GAN instead of original network structure of \cite{alfarraj2019semi} in semi-supervised learning \cite{wu2021semi, meng2020semi}, which leads to a better result. Also, there are many studies that extend semi-supervised learning to other geological fields, such as reservoir properties estimation \cite{di2020semi} and subsurface properties estimation \cite{di2021estimating}.

Uncertainty estimation is also useful in impedance inversion, since it provides a reference for the confidence of the inversion results. Traditional research includes using Markov-chain Monte Carlo (MCMC) to analyse the uncertainty of inverse model \cite{cho2018quasi}. For deep learning algorithms, one way is to embed the existing neural network into a Bayesian framework and to find the posterior distribution \cite{blundell2015weight}. Siahkoohi et al. make use of deep prior to randomly initialize CNN for seismic imaging and uncertainty quantification \cite{siahkoohi2020deep}. Choi et al. introduce variational dropout as a Bayesian approximation for neural network and evaluate prediction uncertainty \cite{choi2020uncertainty}. Ma et al. apply deep evidential regression \cite{amini2019deep} in semi-supervised learning to realize uncertainty estimation \cite{ma2021improved}. However, Deep evidential regression places priors directly over the likelihood function, instead of placing priors on network weights as is usually done in Bayesian neural network. Although deep evidential regression saves the cost of sampling, its neural network is still deterministic and does not provide a convincing estimation of uncertainty. 

In this work, we propose a semi-supervised framework based on Bayesian neural network. Each parameter in the network is assumed to follow a prior of Gaussian, and our aim is to infer the posterior distribution of these parameters. We use pre-training and a local reparameterization trick \cite[]{kingma2015variational} to reduce the computational cost. This framework provides uncertainty estimation for semi-supervised impedance inversion with good interpretability from a Bayesian perspective. The rest of this paper is organized as follows. In section \ref{Related Work}, we introduce the deep learning algorithms used in this article, including original semi-supervised learning and Bayesian neural network. In section \ref{Methodology}, the details of our framework are displayed, especially the strategy to realize uncertainty estimation. Section \ref{Experiments} presents and analyzes the experimental results based on Marmousi2 model and SEAM model. Section \ref{Conclusion} summarizes the whole passage.

\section{Related Work}\label{Related Work}

\subsection{Semi-supervised learning}\label{semi-supervised}
 In practice, it is not easy to obtain the true value of impedance associated with the seismic data, since it needs well drilling wherever impedance information is unknown. Therefore, supervised learning is hard to be exerted unless we use other geological data for training. Semi-supervised learning for impedance inversion is proposed by Motaz Alfarraj and Ghassan AlRegib \cite{alfarraj2019semisupervised,alfarraj2019semi}. It is designed for realizing impedance inversion with seismic data and a small proportion of true labels. 
 
 In semi-supervised learning problem, suppose the dataset is $\mathcal{D}=\{\mathbf{X},\mathbf{Y}\}=\{\mathbf{x}_i,\mathbf{y}_i\}_i$, where $\mathbf{x}_i$ is seismic data of the $i$-th trace, and $\mathbf{y}_i$ is the corresponding AI trace. A subset of AI is assumed to be observed from well logs, which is denoted by $\mathbf{y}_{i'}$, and the corresponding seismic data is denoted by $\mathbf{x}_{i'}$. The network structure of semi-supervised learning resembles the AutoEncoder in artificial intelligence. A parameterized function (neural network) from $\mathbf{X}$ to $\mathbf{Y}$ is called an inverse model, denoted by $f_{\mathbf{w}_1}(\mathbf{x})$. Similarly a forward model from $\mathbf{Y}$ to $\mathbf{X}$ is denoted by $g_{\mathbf{w}_2}(\mathbf{y})$, where $\mathbf{w}_1$ and $\mathbf{w}_2$ are the function (network) parameters.  When training the model, we update $\mathbf{w}_1$ and $\mathbf{w}_2$ to minimize the loss function
\begin{equation}
  L=\alpha_1 \sum_{i'}L_1(f_{\mathbf{w}_1}(\mathbf{x}_{i'}),\mathbf{y}_{i'})+\alpha_2 \sum_{i}L_2(g_{\mathbf{w}_2}(f_{\mathbf{w}_1}(\mathbf{x}_{i})),\mathbf{x}_{i}),\label{loss-semi}
\end{equation}
where $\alpha_1$, $\alpha_2$ are trade-off weights, and $L_1$, $L_2$ are loss functions defined on $\mathbf{x}$-space and $\mathbf{y}$-space. 

Semi-supervised learning is very useful when there is no adequate information about impedance. Nevertheless, this method has several drawbacks. Just as most of the deep learning methods, semi-supervised learning has poor robustness and usually produces artifacts on predicted AI images. 

\subsection{Bayesian neural network} \label{BNN}
Bayesian neural network is a kind of deep learning method which embeds the existing neural network into a Bayesian framework and aims to find the posterior distribution \cite{blundell2015weight}. In typical deep learning methods, neural network can be seen as a function with parameters $\mathbf{w}$. Given the training dataset $\mathcal{D}=\{\mathbf{x}_i,\mathbf{y}_i\}_i$, the parameters $\mathbf{w}$ are determined by maximum likelihood estimation:

\begin{equation}
  \begin{aligned}
  \mathbf{w}^{\mathrm{MLE}}&=\underset{\mathbf{w}}{\mathrm{arg}\max}\ \mathrm{log} P(\mathcal{D}|\mathbf{w})\\
  =&\underset{\mathbf{w}}{\mathrm{arg}\max}\sum_i\mathrm{log} P(\mathbf{y}_i|\mathbf{x}_i,\mathbf{w}),
  \end{aligned}
\end{equation}
where $P(\mathcal{D}|\mathbf{w})$ is the likelihood function.

In Bayesian neural network, the parameters are initially assigned a prior $P(\mathbf{w})$ and we aim to infer the posterior distribution $P(\mathbf{w}|\mathcal{D})$. Computation and inference are therefore more robust under perturbation of the weights. In addition, the output of Bayesian neural network is not deterministic, which provides an uncertainty estimation for the prediction result.

However, it is intractable to solve $P(\mathbf{w}|\mathcal{D})$ directly, especially when the network structure is complicated. We instead use a relatively simple distribution $q(\mathbf{w}|\theta)$ to approximate the real posterior $P(\mathbf{w}|\mathcal{D})$, where $\theta$ are hyperparameters governing $\mathbf{w}$. During the training process, $\theta$ is updated to minimize the KL divergence between $q(\mathbf{w}|\theta)$ and $P(\mathbf{w}|\mathcal{D})$:
\begin{equation}
  \begin{aligned}
    \theta=&\underset{\theta}{\mathrm{arg}\min}\ \mathrm{KL}[q(\mathbf{w}|\theta)||P(\mathbf{w}|\mathcal{D})]\\
    =&\underset{\theta}{\mathrm{arg}\min}\ \mathrm{KL}[q(\mathbf{w}|\theta)||P(\mathbf{w})]-\mathbb{E}_{q(\mathbf{w}|\theta)}[\mathrm{log}P(\mathcal{D}|\mathbf{w})].
    \end{aligned}
\end{equation}
In most case $q(\mathbf{w}|\theta)$ and $P(\mathbf{w})$ are set to be Gaussian, so that the first term $\mathrm{KL}[q(\mathbf{w}|\theta)||P(\mathbf{w})]$ could be calculated analytically.  The second term $\mathbb{E}_{q(\mathbf{w}|\theta)}[\mathrm{log}P(\mathcal{D}|\mathbf{w})]$ could be evaluated by Monte Carlo (MC) method.

There have been many variants of Bayesian neural network. Local reparameterization trick is used in Bayesian neural network to reduce the computational cost of sampling \cite{kingma2015variational}. Dropout is another way to realize Bayesian deep learning, in which the parameters of network follow the Bernoulli distribution \cite{gal2016dropout}. Relevant studies have developed Bayesian version of many other typical deep learning structures, such as RNN \cite{gal2016theoretically} and CNN \cite{gal2015bayesian}.

Despite its advantages such as strong robustness and providing prediction uncertainty, Bayesian neural network also has some drawbacks. In most cases, the prediction result of Bayesian neural network is less accurate than that of non-Bayesian network. In addition, Bayesian neural network  takes much more time and computational resources to be trained compared with non-Bayesian version. 

\section{Methodology}\label{Methodology}
\subsection{Overall workflow}
In this section, we talk about the methodology of constructing semi-supervised Bayesian neural network. We continue to use the notation in section \ref{Related Work}, and assume that the priors of $\mathbf{w}_1$ and $\mathbf{w}_2$ are Gaussian with zero means:
\begin{equation}
  \begin{aligned}
  P(\mathbf{w}_k)=\mathcal{N}(\mathbf{w}_k|0,\sigma_0^2),\ k=1,2.\label{prior}
  \end{aligned}
\end{equation}
In addition, the likelihood functions are Gaussian as well:
\begin{equation}
  \begin{aligned}
    P(\mathbf{Y}|\mathbf{X},\mathbf{w}_1)&=\prod_{i}\mathcal{N}(\mathbf{y}_{i}|f_{\mathbf{w}_1}(\mathbf{x}_{i}), \beta^{-1}),\\
    P(\mathbf{X}|\mathbf{Y},\mathbf{w}_2)&=\prod_i\mathcal{N}(\mathbf{x}_i|g_{\mathbf{w}_2}(\mathbf{y}_i), \beta^{-1}). \label{likelihood}
  \end{aligned}
\end{equation}
where $\beta^{-1}$ is the variance of the likelihood functions. Since it is intractable to solve the Bayesian posterior $P(\mathbf{w}_k|\mathcal{D})$, we need to find a varitional approximation $q(\mathbf{w}_k|\theta_k)$ which is parameterized by $\theta_k$. Suppose $q(\mathbf{w}_k|\theta_k)$ has the form:
\begin{equation}
  \begin{aligned}
    q(\mathbf{w}_k|\theta_k)&=\prod_jq(w_{kj}|\theta_{kj}),\\
    q(w_{kj}|\theta_{kj})&=\mathcal{N}(w_{kj}|\mu_{kj},\sigma_{kj}^2),\label{posterior}
  \end{aligned}
\end{equation}
where $w_{kj}$ represents the $j$th component of $\mathbf{w}_k\ (k=1,2)$. The parameters in the model are thought to be independent \cite{blundell2015weight}. In addition, let
\begin{equation}
  \sigma_{kj}=\mathrm{log}(1+\mathrm{exp}(\rho_{kj})) \label{rho}
\end{equation}
to ensure that $\sigma_{kj}$ remains non-negative \cite{kingma2013auto}. $\rho_{kj}$ is updated instead of $\sigma_{kj}$.

To reduce the computational cost and improve the prediction accuracy, the model is trained in two steps. In the first step, the model is pre-trained in order to determine the value of $\mu_{kj}$. In this step $\mathbf{w}_1, \mathbf{w}_2$ could be viewed as deterministic parameters: $w_{kj}=\mu_{kj}$, and we update $\mu$ by minimizing the loss function as follows:
\begin{equation}
  \hat{\mu}_{1,2}=\underset{\mu_{1,2}}{\mathrm{arg}\min}\ \alpha_1\sum_{i'}\lVert \mathbf{y}_{i'}-f_{\mu_1}(\mathbf{x}_{i'})\rVert^2+\alpha_2\sum_i\lVert \mathbf{x}_i-g_{\mu_2}(f_{\mu_1}(\mathbf{x}_i))\rVert^2. \label{loss}
\end{equation}

In the second step, $\mu_{kj}$ remains unchanged. Suppose $\hat{\mu}_{kj}$ is the mean of $w_{kj}$ obtained by pre-training and $\overline{\theta}_{kj}=(\hat{\mu}_{kj}, \rho_{kj})$. We update $\rho_{kj}$ to minimize the KL divergence between $q(\mathbf{w}_k|\overline{\theta}_k)$ and $P(\mathbf{w}_k|\mathcal{D})$:
\begin{equation}
  \begin{aligned}
    \hat{\rho}_{1,2}&=\underset{\rho_{1,2}}{\mathrm{arg}\min}\sum_{k=1,2}\mathrm{KL}[q(\mathbf{w}_k|\overline{\theta}_k)||P(\mathbf{w}_k|\mathcal{D})]\\
    &=\underset{\rho_{1,2}}{\mathrm{arg}\min}\sum_{k=1,2}\mathrm{KL}[q(\mathbf{w}_k|\overline{\theta}_k)||P(\mathbf{w}_k)]-\mathbb{E}_{q(\mathbf{w}_k|\overline{\theta}_k)}[\mathrm{log}P(\mathcal{D}|\mathbf{w}_k)],\label{KL}
  \end{aligned}
\end{equation} 
where
\begin{equation}
  \begin{aligned}
    q(\mathbf{w}_k|\overline{\theta}_k)=\prod_j\mathcal{N}(w_{kj}|\hat{\mu}_{kj}, \sigma_{kj}^2).
  \end{aligned}
\end{equation}

In the end, the approximated posterior distribution $q(\mathbf{w}_k|\hat{\theta}_k)$, where $\hat{\theta}_k=(\hat{\mu}_k, \hat{\rho}_k)$, is obtained. For a new value $\hat{\mathbf{x}}$, the predictive distribution of impedance $\hat{\mathbf{y}}$ can be calculated as:
\begin{equation}
  \begin{aligned}
    p(\hat{\mathbf{y}}|\hat{\mathbf{x}},\mathbf{w}_1,\mathbf{X},\mathbf{Y})=\int q(\mathbf{w}_1|\hat{\theta}_1)P(\hat{\mathbf{y}}|\hat{\mathbf{x}},\mathbf{w}_1)\mathrm{d}\mathbf{w}_1.
  \end{aligned} \label{prediction}
\end{equation}

\begin{figure}[!p]
  \centering
  \setlength{\abovecaptionskip}{2cm}
  \includegraphics[width = 0.8\textwidth]{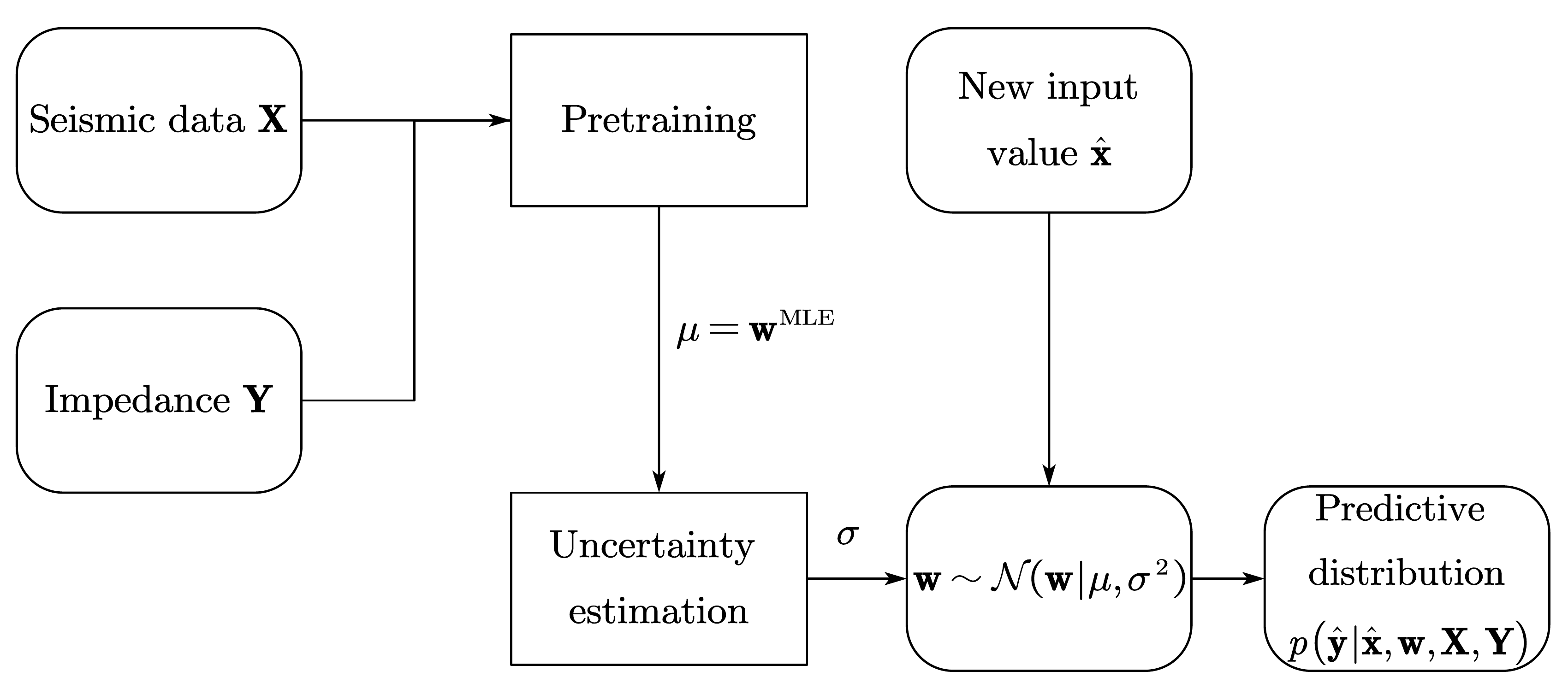}
  \caption{Overall workflow of the proposed method.}\label{workflow}
\end{figure}

The mean and variance of this integral could be evaluated by Monte Carlo sampling. The overall workflow of the proposed method is shown in Figure \ref{workflow}.

\subsection{Pre-training}
In pre-training, we assume that parameters $\mathbf{w}$ in Bayesian neural network are deterministic. In other words, each parameter $w_{kj}$ follows a Gaussian with mean $\mu_{kj}$ and variance $0$, and we minimize the loss in equation \eqref{loss} with respect to $\mu_{kj}$. 

\begin{figure}[!p]
  \setlength{\abovecaptionskip}{1cm}
	\begin{center}
    \subfigure[]{
      \includegraphics[width=12cm]{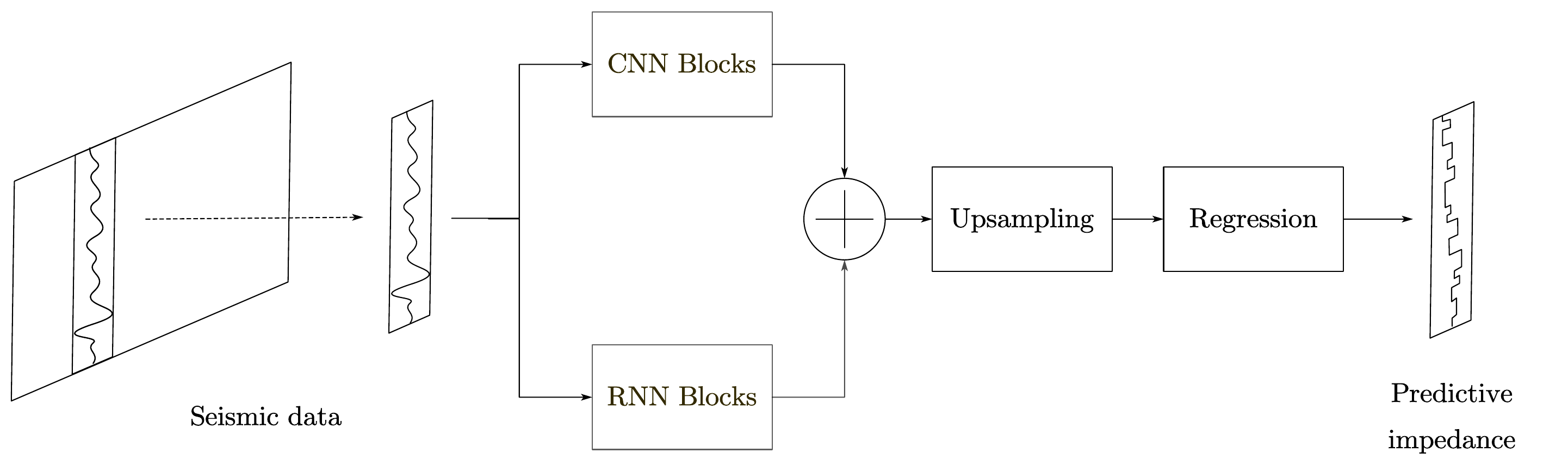}
    } 
    \subfigure[]{
      \includegraphics[width=12cm]{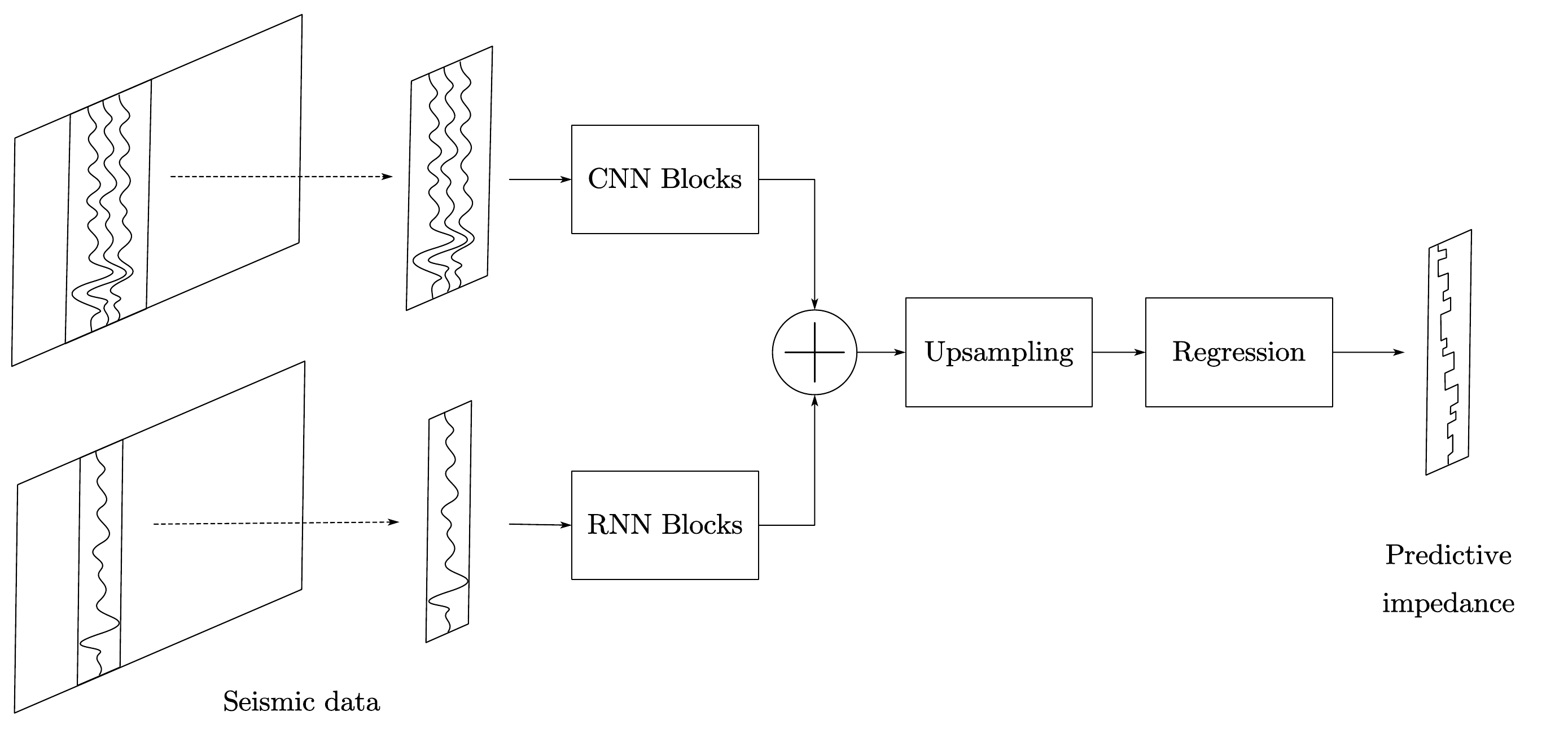} 
    }
	  \caption{Comparision of two structures of inverse model: \textbf{(a)} structure of inverse model in \citet{alfarraj2019semi}; \textbf{(b)} proposed structure.}  \label{Semi}
  \end{center} 
\end{figure}

Our pre-training is based on the semi-supervised learning network structure proposed by \citet{alfarraj2019semi}. The inverse model of the network is composed of four submodules: sequence modeling, local pattern analysis, upsampling and regression. Sequence modeling consists of three layers of Gate Recurrent Unit (GRU). Local pattern analysis consists of three parallel layers of 1-$d$ CNN with different dilations and subsequent three layes of 1-$d$ CNN with different kernel sizes. There is also a group norm layer between two connected CNN layers. The input seismic is processed by these two submodules at the same time to capture the data features. The two outputs from CNN layers and GRU layers are added up and sent to upsampling, which is to ensure the final outputs and the labels having the same dimension by using two deconvolution layers. In the end, regression submodules map the unscaled data from features space to the target space, using a  GRU layer and a linear layer. The structure of the inverse model is shown in Figure \ref{Semi}(a). The forward model of this network contains a 2-layer CNN to calculate seismic from AI. The network achieves good performance even when there are less than $1\%$ of AI labels are observed.

\begin{figure}[!p]
  \centering
  \setlength{\abovecaptionskip}{1cm}
  \includegraphics[width = 0.9\textwidth]{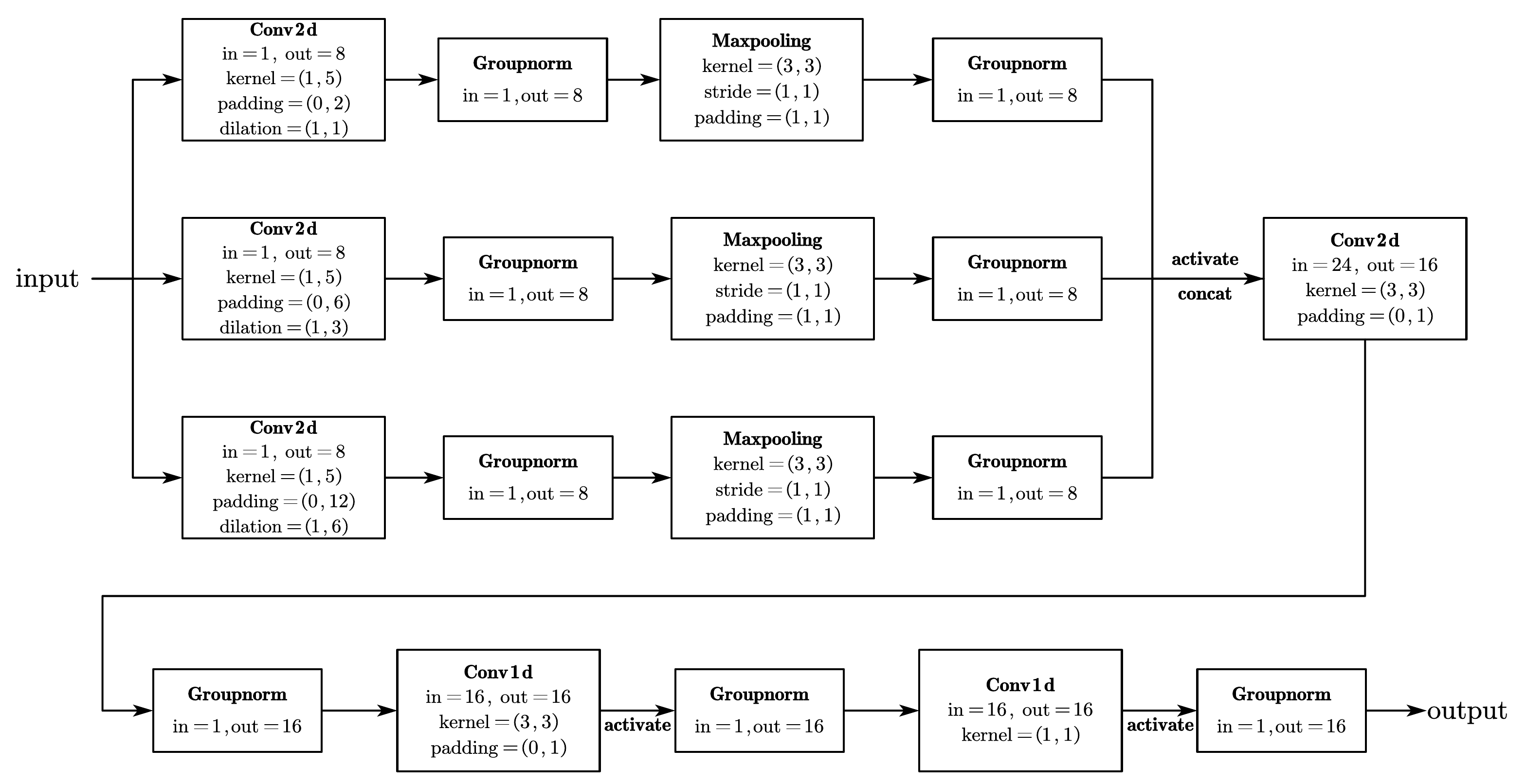}
  \caption{Structure of CNN layers in the proposed inverse model.}\label{CNN blocks}
\end{figure}

The original semi-supervised learning algorithm is by nature a trace-by-trace prediction method, which may result in lateral discontinuities as shown in Figure \ref{pre-Marf}(e). To alleviate this problem, we merge the network structure proposed by \citet{mustafa2020spatiotemporal} into the inverse model of \citet{alfarraj2019semi}. First, for a certain trace of seismic data $\mathbf{x}_i\in \mathbb{R}^{T\times 1}\ (i=1,\cdots,N)$ and a positive integer $h>0$, we collect its adjacent traces $\mathbf{x}_{i-h},\cdots,\mathbf{x}_{i+h}$ and concatenate them together to form a data matrix $\mathbf{X}_i$ with a shape $(T,2h+1)$. On the boundaries, let $\mathbf{x}_{i-j}=\mathbf{0}$ if $j\geqslant i$ and $\mathbf{x}_{i+j}=\mathbf{0}$ if $j\geqslant N-i+1$. Next, data matrix $\mathbf{X}_i$ is input into the CNN layers of inverse model. The CNN layers is modified by changing the three parallel 1-$d$ convolutions and the first subsequent 1-$d$ convolution with 2-$d$ convolutions. Besides, we add a 2-$d$ maxpooling layer after every convolutional operations. GRU layers in the model is designed to capture the sequential relationship of seismic data. Therefore, the data is input into GRU layers trace by trace. The proposed structure of inverse model is shown in Figure \ref{Semi}(b), and the detailed structure of CNN layers in the proposed inverse model is shown in Figure \ref{CNN blocks}.

\subsection{Uncertainty estimation}\label{Uncertainty}
In uncertainty estimation, we update $\rho$ to minimize the right side of equation \eqref{KL}. Since the form of the prior, likelihood and the approximated posterior have been defined in equation \eqref{prior}, \eqref{likelihood} and \eqref{posterior}, the right side of \eqref{KL} can be calculated by
\begin{equation}
  \begin{aligned}
    \mathrm{KL}[q(\mathbf{w}_k|\overline{\theta}_k)||P(\mathbf{w}_k)]=-\frac{1}{2}\sum_j \{\mathrm  {log}\sigma_{kj}^2-\frac{\sigma_{kj}^2}{\sigma_{0}^2}\}+\mathrm{const} \label{KL2}
  \end{aligned}
\end{equation}
and
\begin{equation}
  \begin{aligned}
    &\mathbb{E}_{q(\mathbf{w}_k|\overline{\theta}_k)}[\mathrm{log}P(\mathcal{D}|\mathbf{w}_k)]\\[2mm]
    &=-\frac{1}{\beta}\mathbb{E}_{q(\mathbf{w}_k|\overline{\theta}_k)}\{\sum_{i'}\lVert \mathbf{y}_ {i'}-f_{\mathbf{w}_1}(\mathbf{x}_{i'})\rVert^2+\sum_i\lVert \mathbf{x}_i-g_{\mathbf{w}_2}(f_ {\mathbf{w}_1}(\mathbf{x}_i))\rVert^2\}+\mathrm{const}\\[-2mm]
    &\approx-\frac{1}{\beta M}\sum_{m=1}^M\{\sum_{i'}\lVert \mathbf{y}_{i'}-f_{\mathbf{w}_1^{(m)}}  (\mathbf{x}_{i'})\rVert^2+\sum_i\lVert \mathbf{x}_i-g_{\mathbf{w}_2^{(m)}}(f_{\mathbf{w}_1^{(m) }}(\mathbf{x}_i))\rVert^2\}+\mathrm{const}. \label{expectation}
\end{aligned}
\end{equation}
The final step uses Monte Carlo approximation to evaluate the expectation, where $\mathbf{w}_1^{(m)}$ and $\mathbf{w}_2^{(m)}$ are the $m$-th sample drawn from $q(\mathbf{w}_1|\overline{\theta}_1)$ and $q(\mathbf{w}_2|\overline{\theta}_2)$. Therefore, the objective function is:
\begin{equation}
  \begin{aligned}
    \mathcal{F}(\rho,\mathcal{D})=&-\frac{1}{\beta M}\sum_{m=1}^M\{\sum_{i'}\lVert \mathbf{y}_{i'}  -f_{\mathbf{w}_1^{(n)}}(\mathbf{x}_{i'})\rVert^2+\sum_i\lVert \mathbf{x}_i-g_{\mathbf{w}_2^{(n) }}(f_{\mathbf{w}_1^{(n)}}(\mathbf{x}_i))\rVert^2\}\\ &-\frac{1}{2}\sum_{k=1,2}\sum_j \{\mathrm {log}\sigma_{kj}^2-\frac{\sigma_{kj}^2}{\sigma_{0}^2}\}.\label{objective}
  \end{aligned} 
\end{equation}

In practice, we apply the local reparameterization trick instead of sampling the Gaussian weights and bias to calculate the objective function \eqref{objective}. Consider a linear operation
\begin{equation}
  \begin{aligned}
    \mathbf{Y}=\mathbf{XW}+\mathbf{b},\label{linear}
  \end{aligned}
\end{equation}
where $\mathbf{W}$ and $\mathbf{b}$ are parameters. Suppose for any $w_{ij}\in \mathbf{W}, b_i\in \mathbf{b}$, 
\begin{equation}
  \begin{aligned}
    w_{ij}\sim \mathcal{N}(w_{ij}|\mu_{ij},\sigma_{ij}^2),\ b_i\sim \mathcal{N}(b_i|\mu'_{i}, {\sigma'_{i}}^2),\label{wb}
  \end{aligned}
\end{equation}
then for any $y_{mj}\in \mathbf{Y}$,
\begin{equation}
  \begin{aligned}
    y_{mj}\sim \mathcal{N}(y_{mj}|\gamma_{ij},\delta_{ij}^2),\label{y}
  \end{aligned}  
\end{equation}
where
\begin{equation}
  \begin{aligned}
    \gamma_{mj}=\sum_ix_{mi}\mu_{ij}+\mu'_i,\ \delta_{mj}^2=\sum_ix_{mi}^2\sigma_{ij}^2+{\sigma'_i}^2.\label{reparameterization}
  \end{aligned}
\end{equation}

In the inverse model, the convolution layers, deconvolution layers and group norm layers could be seen as linear operations as in \eqref{linear} and \eqref{wb}. Before uncertainty estimation, we have fixed the value of $\mu$ and $\mu'$. So we can get the objective function as follows. First, for each layer, we initialize the hyperparameters $\rho$ and $\rho'$ for every weights and biases, and calculate the KL divergence between the approximated posterior and prior of this layer by \eqref{KL2}. Second, calculate the value of $\gamma$ and $\delta$ by \eqref{reparameterization}, where $\mu_{ij}=\hat{\mu}_{ij}$ and $\mu'_{ij}=\hat{\mu'}_{ij}$ are from pre-training. Third, sample a standard Gaussian random variable $\epsilon_{mj}\sim \mathcal{N}(0,1)$ and calculate the output $\mathbf{Y}$ by
\begin{equation}
  \begin{aligned}
    y_{mj}=\gamma_{mj}+\delta_{mj}\epsilon_{mj},\ \forall y_{mj}\in \mathbf{Y}. \label{gety}
  \end{aligned}
\end{equation}

The GRU layer in the inverse model, denoted by $f_{\hat{\mu}}(\mathbf{X})$, is a non-linear operation. Nevertheless it could be approximated by a linear operation:
\[
  \begin{aligned}
    f_{\hat{\mu}}(\mathbf{X})\approx \mathbf{X}\widetilde{\mathbf{W}}+\widetilde{\mathbf{b}}.
  \end{aligned}
\]
So that we construct another two parameters $\widetilde{\mathbf{W}}$ and $\widetilde{\mathbf{b}}$ of form \eqref{wb} for each GRU layer. Then get the corresponding $\delta_{mj}$ by \eqref{reparameterization}
and compute the output:
\begin{equation}
  \begin{aligned}
    y_{mj}=f_{mj}+\delta_{mj}\epsilon_{mj},\ \forall y_{mj}\in \mathbf{Y}, \label{gety2}
  \end{aligned}
\end{equation}
where $f_{mj}$ is the $(m,j)$ element of $f_{\hat{\mu}}(\mathbf{X})$.

Finally, after the output of the last layer is obtained, the expectation is calculated by \eqref{expectation}. We add up the KL divergence from each layer, together with the result in \eqref{expectation} to get the result of $\mathcal{F}(\rho,\mathcal{D})$. After that, it needs to update hyperparameters $\rho$ by minimizing $\mathcal{F}(\rho, \mathcal{D})$. Then $\rho$ is optimized iteratively until convergence. The complete computational process is shown in Algorithm \ref{algorithm}. Note that there is no GRU layers in forward model.

\begin{algorithm}[ht]
  \setstretch{1.05}
  \caption{Algorithm for updating hyperparameters $\rho$} \label{algorithm}
  \hspace*{0.02in} {\bf Input:}
  $\mu$ and $\mu'$ for every weights and biases in each layer, prior variance $\sigma_0^2$, seismic data $\mathbf{X}_0$, 
  \hspace*{0.02in} observed AI $\mathbf{Y}_0'$.\\
  \hspace*{0.02in} {\bf Output:} 
  updated $\rho$ and $\rho'$ for every weights and biases in each layer.
  \begin{algorithmic}[1]
    \State Initialize $\rho$ for and $\rho'$ every weights and biases in each layer (for $\widetilde{\mathbf{W}}$ and $\widetilde{\mathbf{b}}$ in GRU layers).
    \While{$\rho$ and $\rho'$ do not converge}
      \State KL=$0$, $E=0$, $\mathcal{F}=0$, $\mathbf{X}=\mathbf{X}_0$, $\mathbf{Y}=\mathbf{0}$, $\mathbf{X}_{\mathrm{pred}}=\mathbf{0}$, $\mathbf{Y}_{\mathrm{pred}}=\mathbf{0}$.
      \For{each layer in inverse model}
        \State $\sigma_{ij}\gets\mathrm{log}(1+\mathrm{exp}(\rho_{ij}))$, ${\sigma_i}'\gets\mathrm{log}(1+\mathrm{exp}(\rho'_i))$.
        \State $\mathrm{KL}\gets\mathrm{KL}-\sum_{i,j}\{\mathrm{log}\sigma_{ij}-\sigma_{ij}^2/(2\sigma_{0}^2)\}-\sum_i \{\mathrm{log}\sigma'_{i}-{\sigma'_{i}}'^2/(2\sigma_{0}^2)\}$.
        \If{not a GRU layer}
          \State $\gamma_{mj}\gets\sum_ix_{mi}\mu_{ij}+\mu'_i$;
        \Else
          \State $\gamma_{mj}\gets f_{mj}$.
        \EndIf
        \State $\delta_{mj}^2\gets\sum_ix_{mi}^2\sigma_{ij}^2+{\sigma'_{i}}^2$ .
        \State Sample standard Gaussians $\epsilon_{mj}\sim \mathcal{N}(0,1)$.
        \State $y_{mj}\gets \gamma_{mj}+\delta_{mj}\epsilon_{mj}$.
        \If{not in the last layer}
          \State $\mathbf{X}\gets\mathbf{Y}$;
        \Else
          \State $\mathbf{Y}_{\mathrm{pred}}\gets\mathbf{Y}$.
        \EndIf
      \EndFor   
      \For{each layer in forward model}
        \State $\sigma_{ij}\gets\mathrm{log}(1+\mathrm{exp}(\rho_{ij}))$, ${\sigma'_{i}}\gets\mathrm{log}(1+\mathrm{exp}(\rho'_i))$.
        \State $\mathrm{KL}\gets\mathrm{KL}-\sum_{i,j}\{\mathrm{log}\sigma_{ij}-\sigma_{ij}^2/(2\sigma_{0}^2)\}-\sum_i \{\mathrm{log}\sigma'_{i}-{\sigma'_{i}}^2/(2\sigma_{0}^2)\}$.
        \State $\gamma_{mj}\gets\sum_iy_{mi}\mu_{ij}+\mu'_i$.
        \State $\delta_{mj}^2\gets\sum_iy_{mi}^2\sigma_{ij}^2+{\sigma'_{i}}^2$ .
        \State Sample standard Gaussians $\epsilon_{mj}\sim \mathcal{N}(0,1)$.
        \State $x_{mj}\gets \gamma_{mj}+\delta_{mj}\epsilon_{mj}$.
        \If{not in the last layer}
          \State $\mathbf{Y}\gets\mathbf{X}$;
        \Else
          \State $\mathbf{X}_{\mathrm{pred}}\gets\mathbf{X}$.
        \EndIf
      \EndFor
      \State Calculate the expectation $E$ in \eqref{expectation} using $\mathbf{X}_{\mathrm{pred}}$, $\mathbf{Y}_{\mathrm{pred}}$, $\mathbf{X}_0$ and $\mathbf{Y}_0'$. 
      \State $\mathcal{F}\gets \mathrm{KL}+E$.
      \State Update $\rho$ and $\rho'$ in order to minimize $\mathcal{F}$.
    \EndWhile
    \State \Return $\rho$ and $\rho'$.
  \end{algorithmic}
\end{algorithm}

Since it is intractable to calculate the integral in \eqref{prediction}, We evaluate the mean and the variance of predictive distribution by Monte Carlo method. It only needs to make several predictions by the inverse model and calculate the sample mean and variance:
\begin{equation}
  \begin{aligned}
  \mathbb{E}(\hat{\mathbf{y}}|\hat{\mathbf{x}},\mathbf{w}_1,\mathbf{X},\mathbf{Y})&\approx \frac{1}{N}\sum_{n=1}^N f_{\mathbf{w}_1^{(n)}}(\hat{\mathbf{x}})\\
  \mathrm{var}(\hat{\mathbf{y}}|\hat{\mathbf{x}},\mathbf{w}_1,\mathbf{X},\mathbf{Y})&\approx \frac{1}{N}\sum_{n=1}^N (f_{\mathbf{w}_1^{(n)}}(\hat{\mathbf{x}})-\frac{1}{N}\sum_{n'=1}^N f_{\mathbf{w}_1^{(n')}}(\hat{\mathbf{x}}))^2 \label{predictionMC}
  \end{aligned}
\end{equation}
where $\mathbf{w}_1^{(n)}\sim q(\mathbf{w}_1|\hat{\theta}_1)$ is the $n$-th sample of $\mathbf{w}_1$. In practice, we just use the result of the pre-training as the sample mean, since the means of parameters in the model are unchanged during uncertainty estimation.

\section{Experiments}\label{Experiments}
\subsection{Marmousi2 model}\label{Marmousi2 model}

In this part, the proposed strategy is tested on the Marmousi2 model \cite{martin2002marmousi}. The data is generated by the open source code of \citet{alfarraj2019semi}. It contains seismic data profile of 2721 traces, each has 470 time samples, and the corresponding AI has 1880 time samples. 

First, We carry out the pre-training on Marmousi2 model. As in \citet{alfarraj2019semi}, we choose 20 traces of AI as observed labels, evenly distributed on the data field. The loss function in \eqref{loss} is set to be 
\begin{equation}
  \begin{aligned}
    L=\frac{1}{N_l} \sum_{i'}L_1(f_{\mu_1}(\mathbf{x}_{i'}),\mathbf{y}_{i'})+\frac{1}{5N_u} \sum_{i}L_2(g_{\mu_2}(f_{\mu_1}(\mathbf{x}_{i})),\mathbf{x}_{i}),\label{loss-Mar}
  \end{aligned}
\end{equation}
where $N_l=20$ is the number of data points whose labels are observed, and $N_u=2721$ is the total number of data points. We train the inverse model and the forward model for 1000 epochs, and use the inverse model to get the approximated mean posterior prediction of the AI.

For comparison, we reproduce the results of \citet{alfarraj2019semi} and \citet{blundell2015weight}. Instead of first pre-training and then estimating uncertainty, \citet{blundell2015weight} updates the mean $\mu$ and the variance $\sigma^2$ of the approximated posterior distribution $\mathcal{N}(\mathbf{w}|\mu,\sigma^2)$ simultaneously. We use the Pytorch package Blitz \cite{esposito2020blitzbdl} to bulid the Bayesian neural network in \citet{blundell2015weight}.

\begin{figure}[!p]
  \setlength{\abovecaptionskip}{0.5cm}
	\begin{center}
    \subfigure[]{
      \includegraphics[width=0.48\textwidth,trim=80 100 180 130,clip]{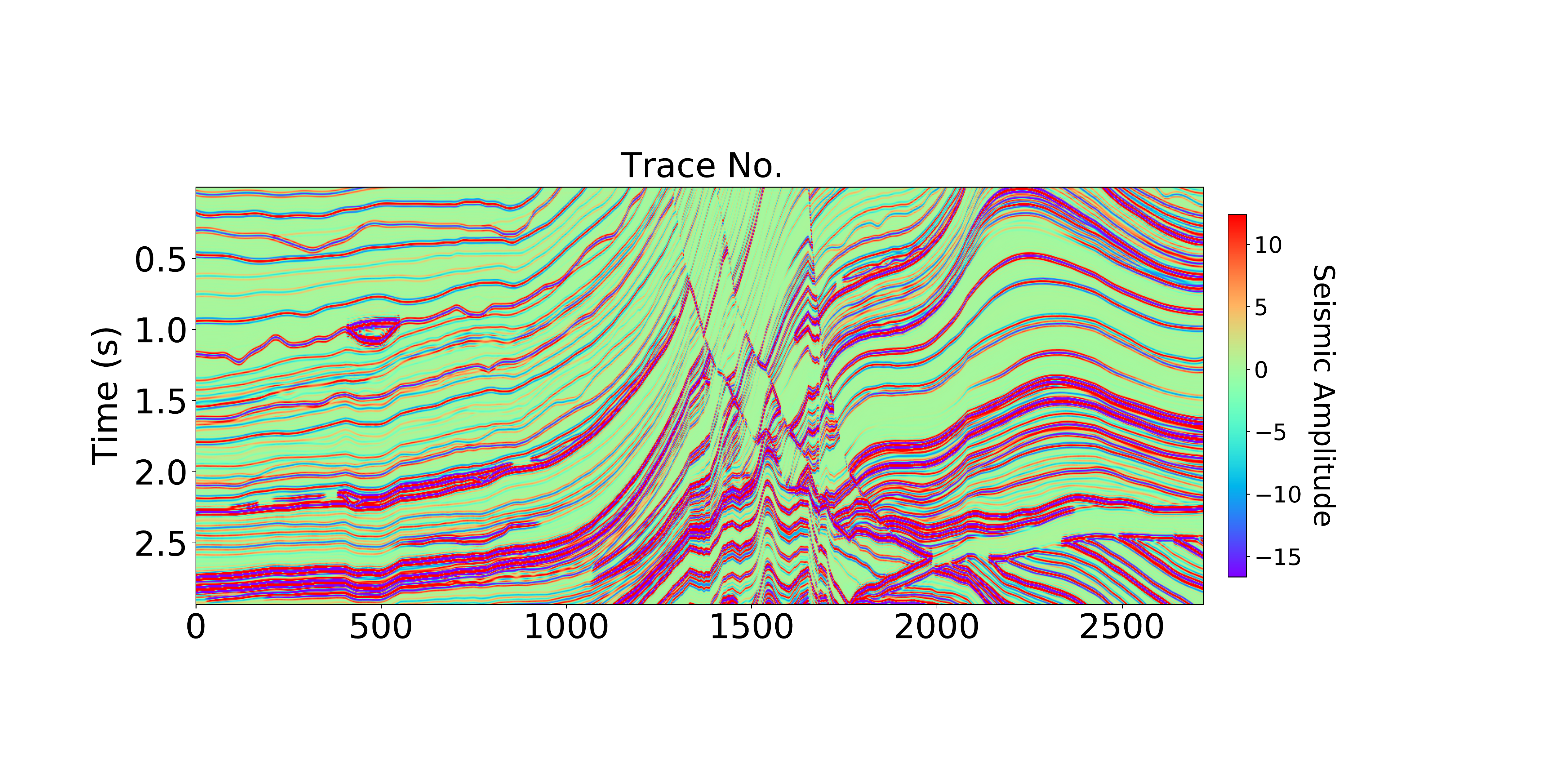} 
    } 
    \subfigure[]{
      \includegraphics[width=0.48\textwidth,trim=80 100 180 130,clip]{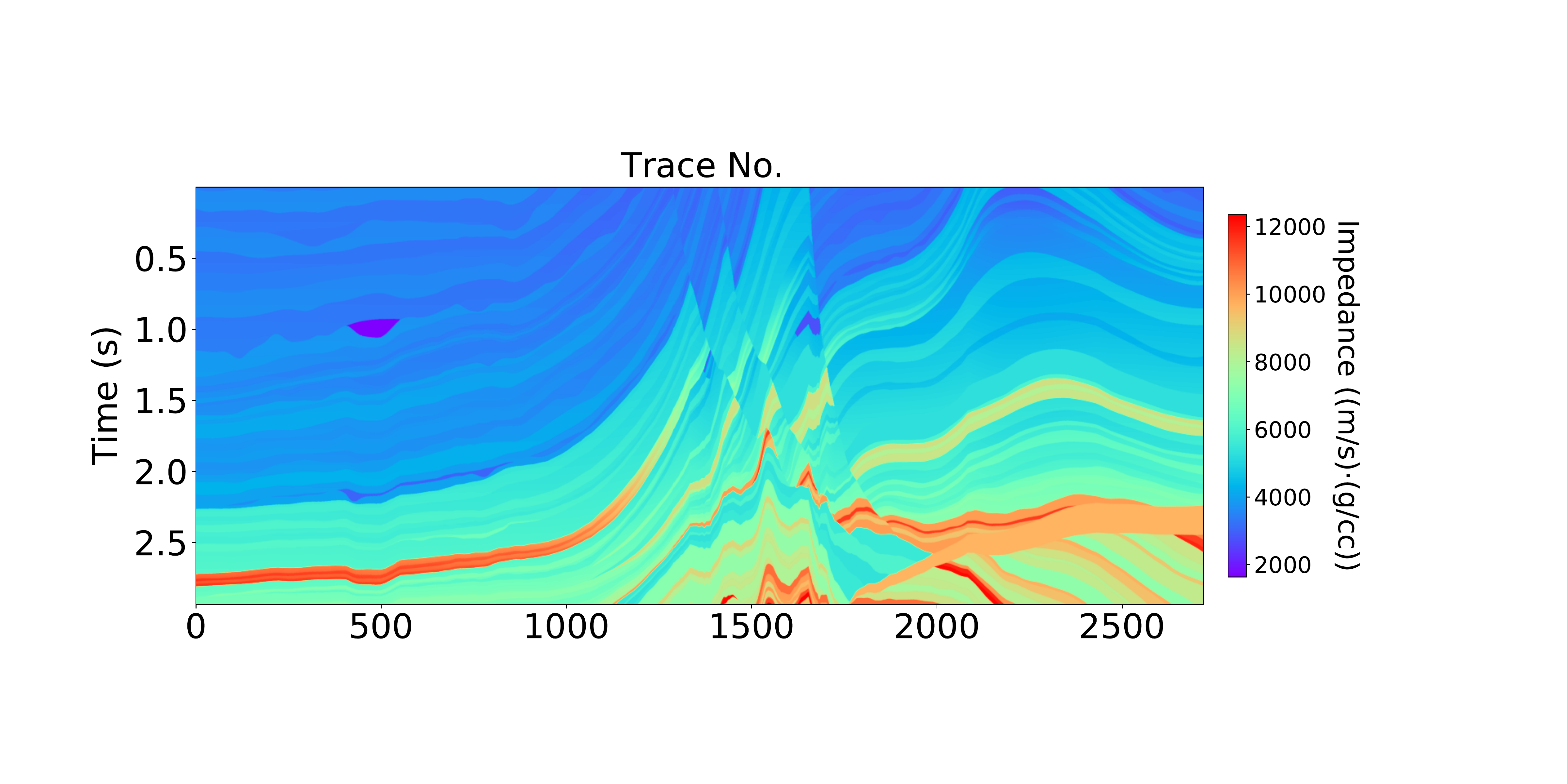}
    }
    \subfigure[]{
      \includegraphics[width=0.48\textwidth,trim=80 100 180 130,clip]{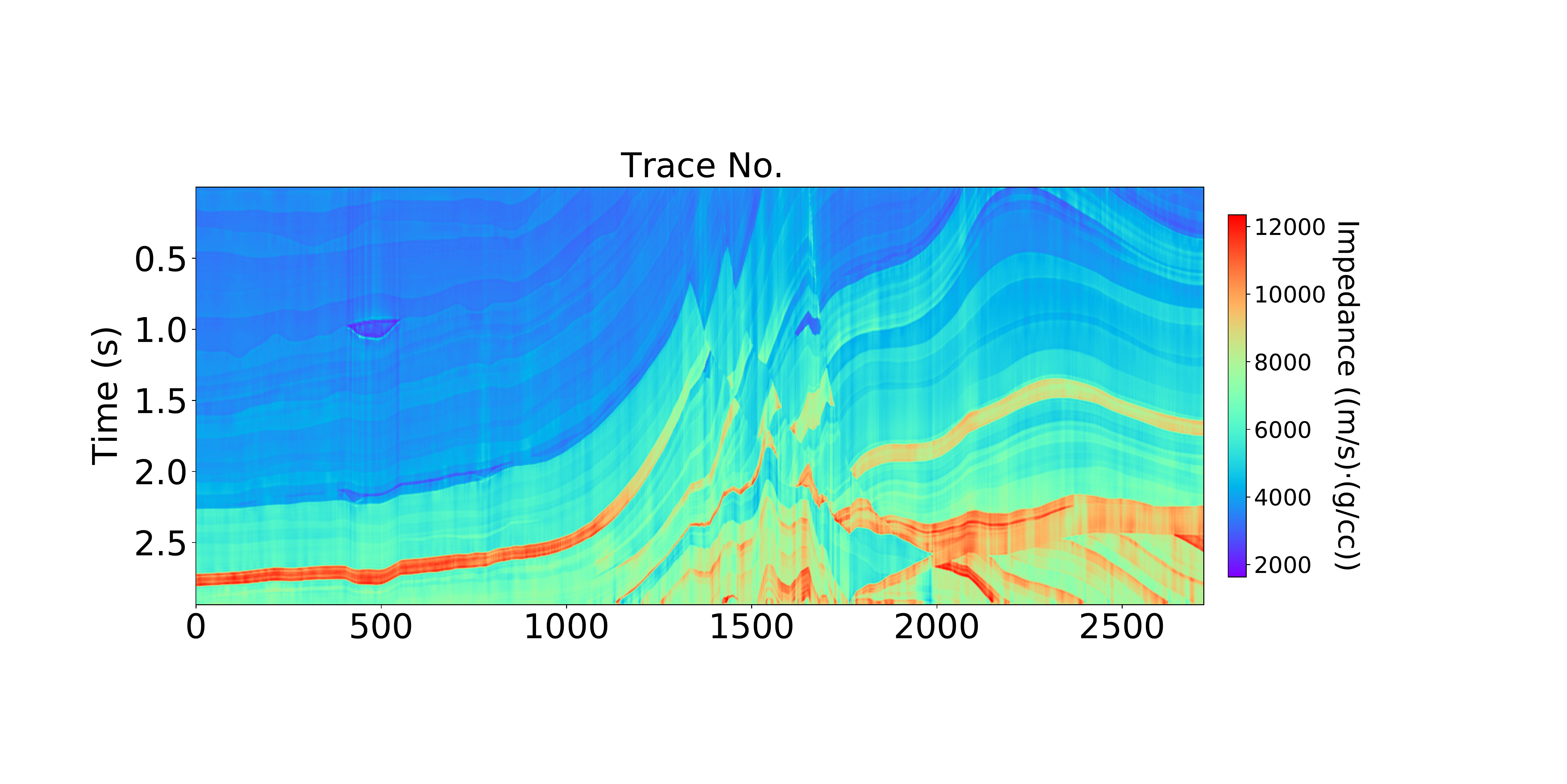}
    }
    \subfigure[]{
      \includegraphics[width=0.48\textwidth,trim=80 100 180 130,clip]{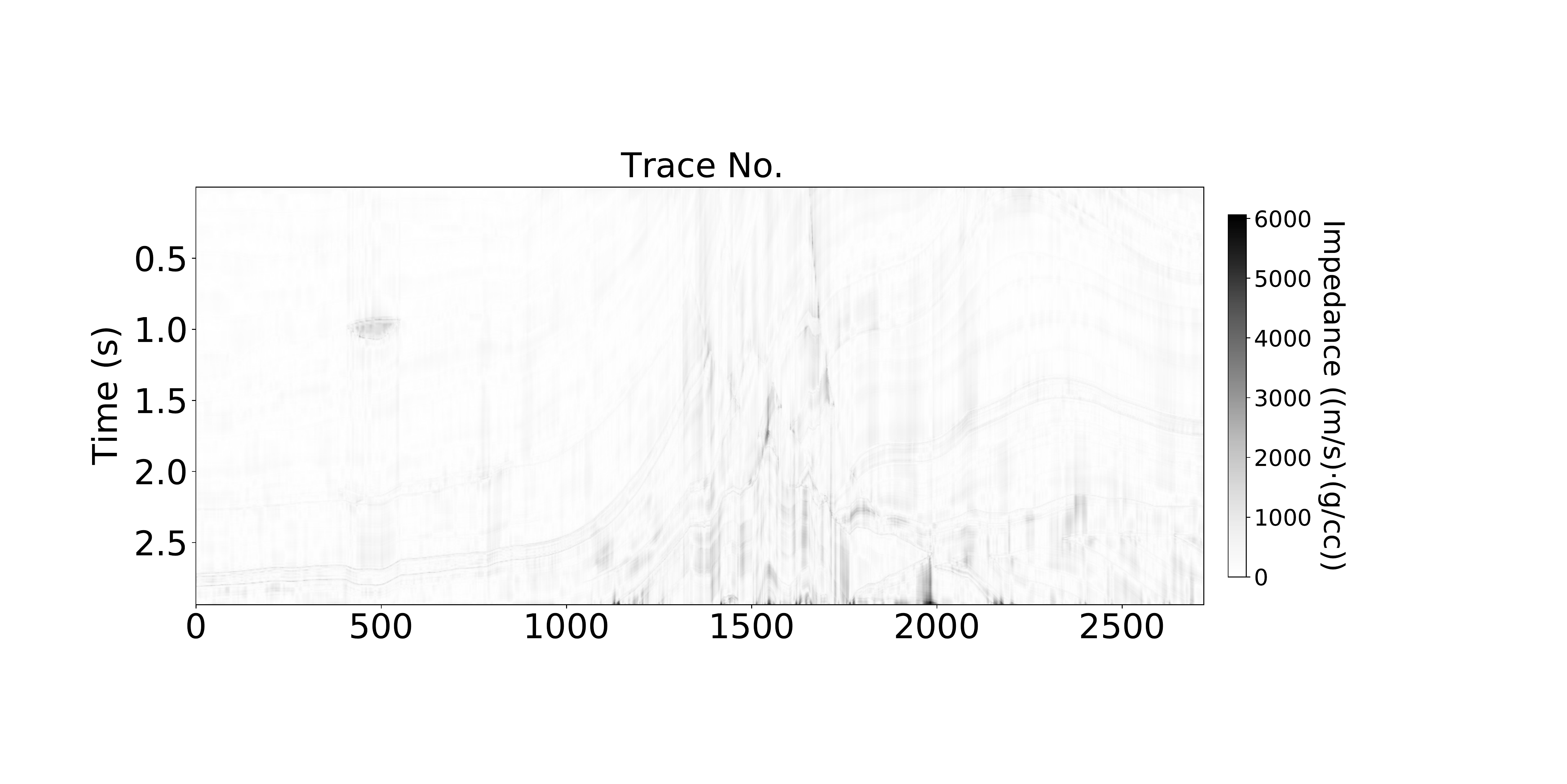}
    }
    \subfigure[]{
      \includegraphics[width=0.48\textwidth,trim=80 100 180 130,clip]{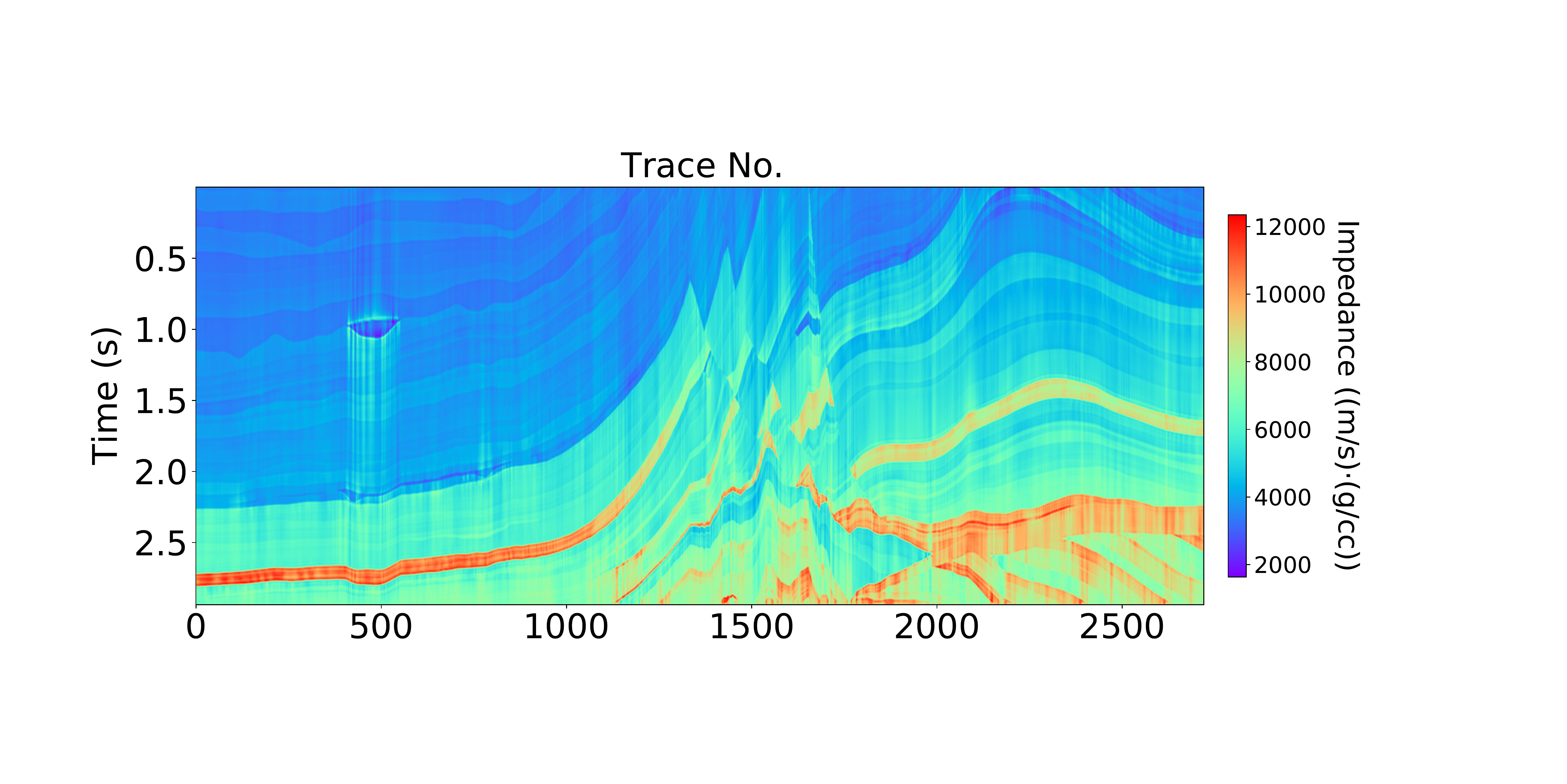}
    }
    \subfigure[]{
      \includegraphics[width=0.48\textwidth,trim=80 100 180 130,clip]{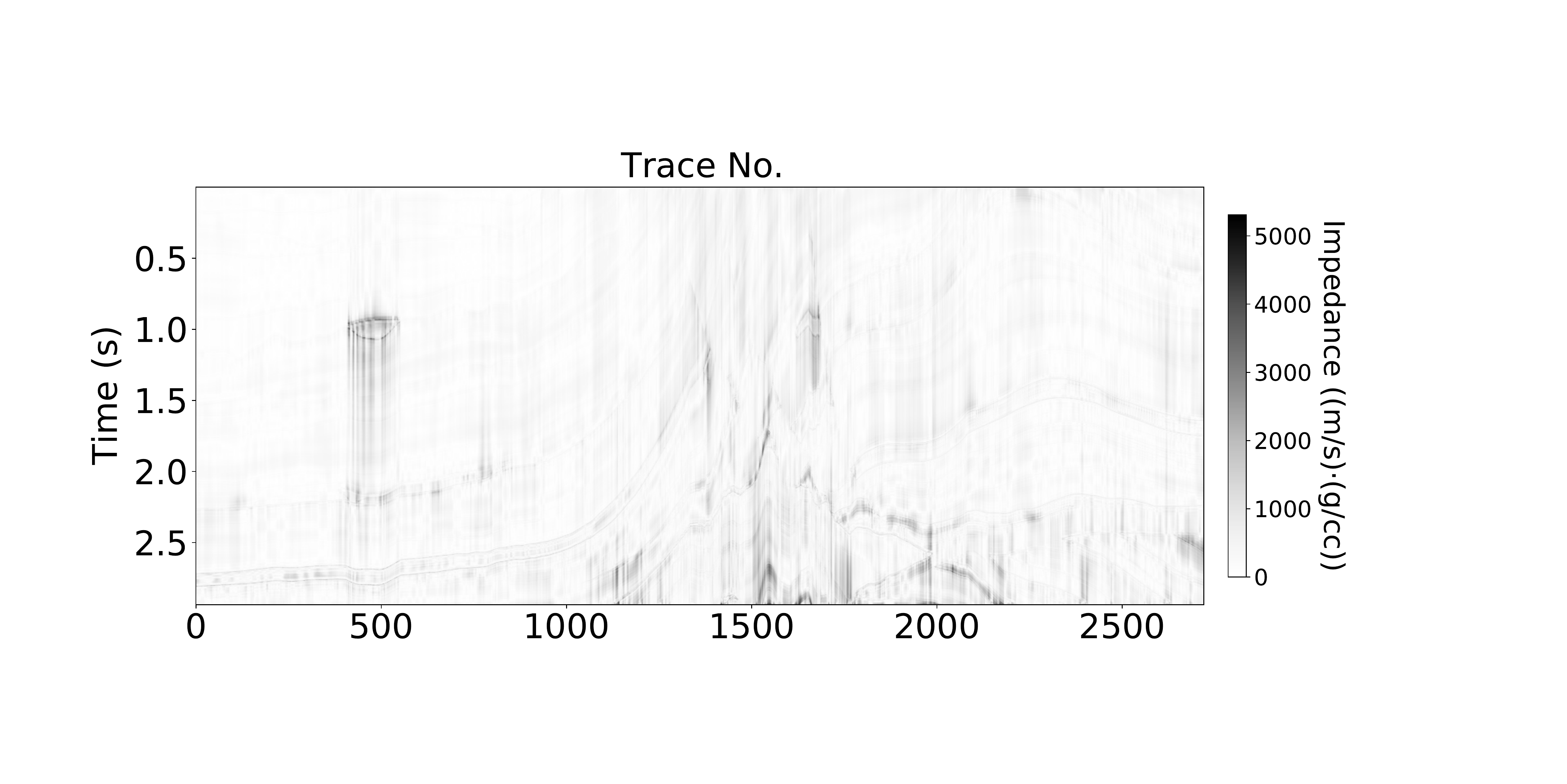}
    }
    \subfigure[]{
      \includegraphics[width=0.48\textwidth,trim=80 100 180 130,clip]{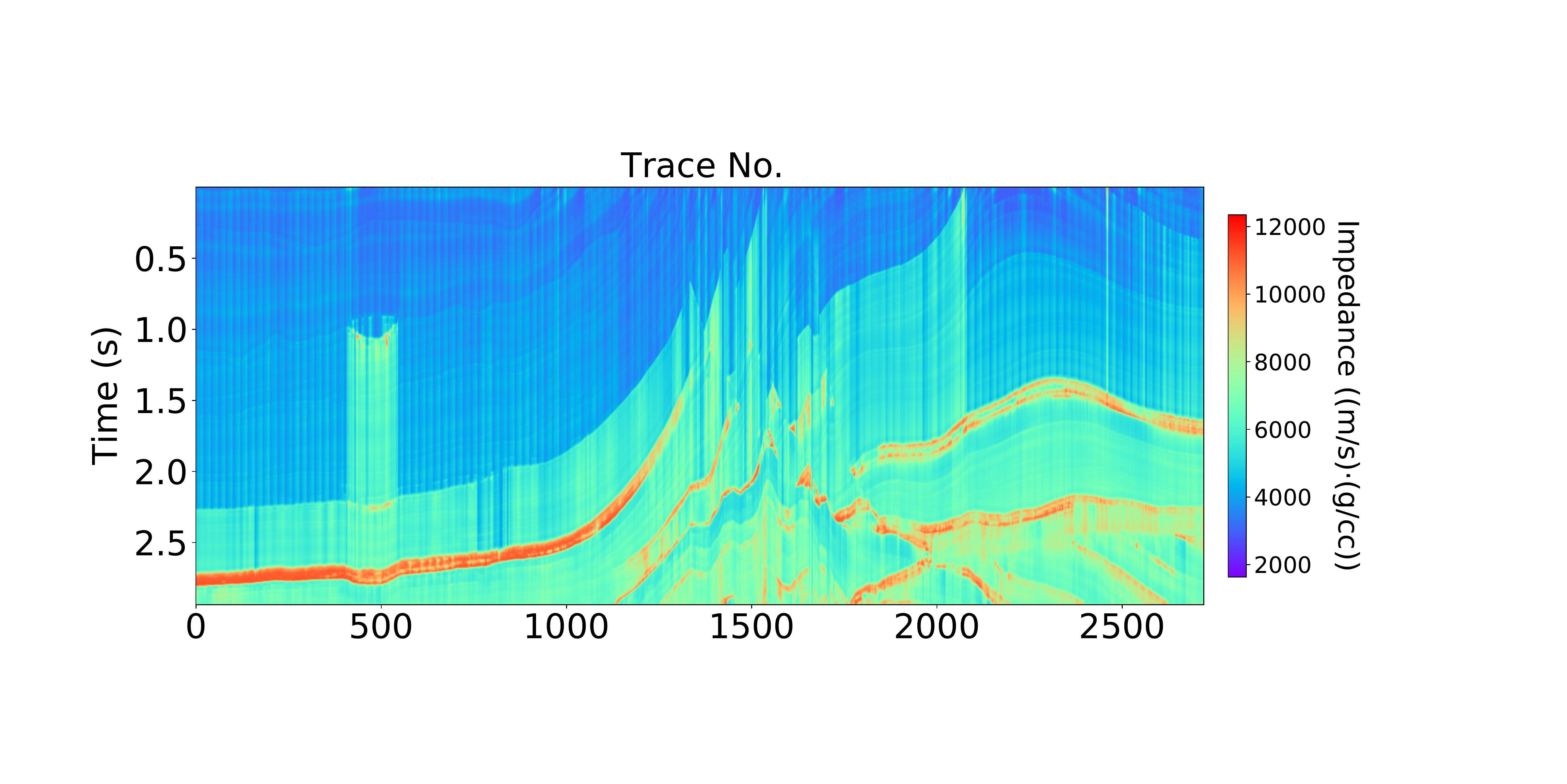}
    }
    \subfigure[]{
      \includegraphics[width=0.48\textwidth,trim=80 100 180 130,clip]{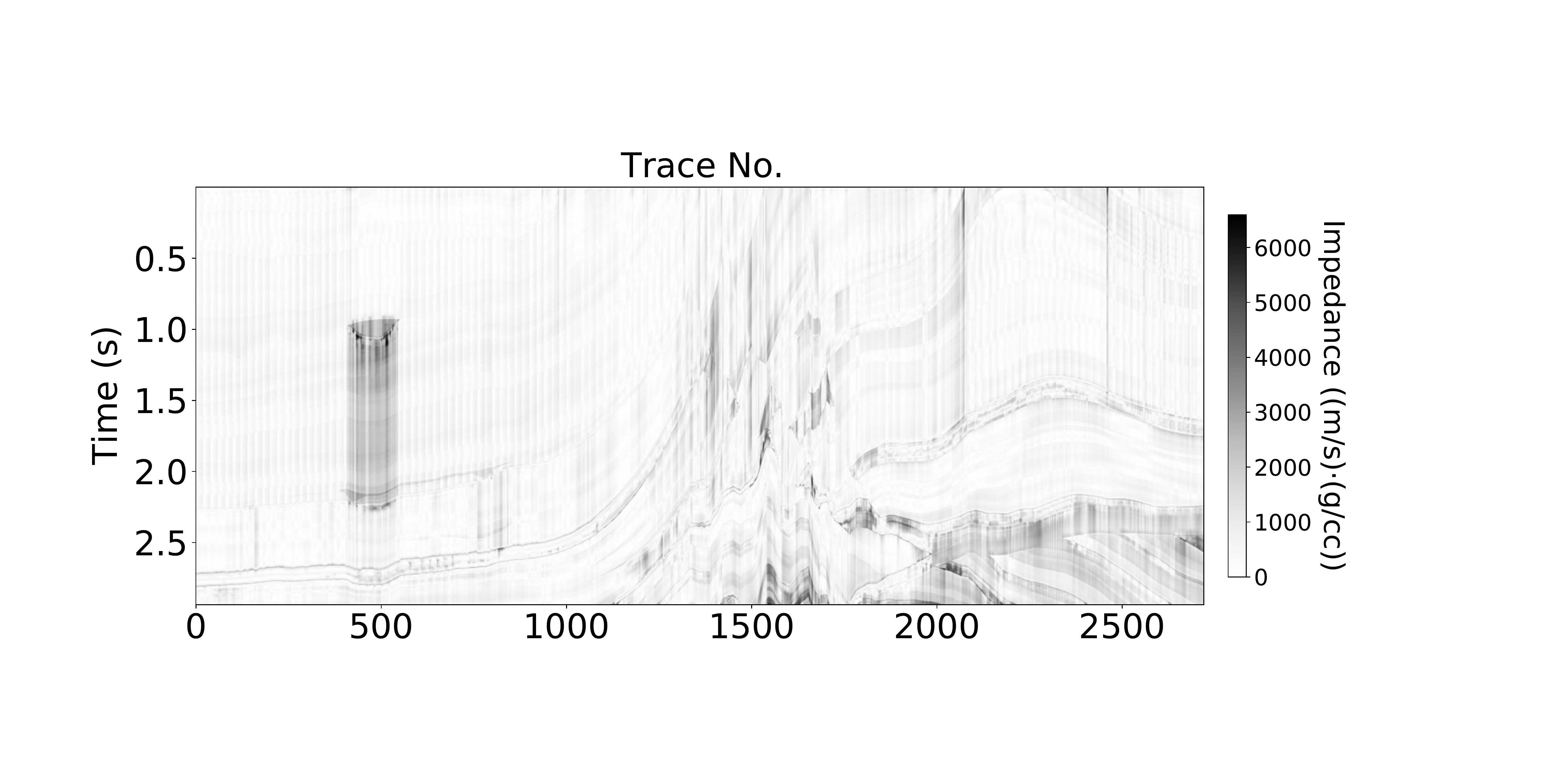}
    }
	  \caption{Seismic data, true AI, predicted AI and absolute difference of the Marmousi2 model test: (\textbf{a}) seismic data for inversion; (\textbf{b}) true AI; (\textbf{c}) AI predicted by proposed method; (\textbf{d}) the absolute difference between (\textbf{b}, \textbf{c}); (\textbf{e}) AI predicted by \citet{alfarraj2019semi}; (\textbf{f}) the absolute difference between (\textbf{b}, \textbf{e});   (\textbf{g}) AI predicted by \citet{blundell2015weight}; (\textbf{d}) the absolute difference between (\textbf{b}, \textbf{g}).} \label{pre-Marf}
  \end{center}
\end{figure}

In Figures \ref{pre-Marf}(a) and \ref{pre-Marf}(b) we show the seismic data for inversion and the target true AI. The mean posterior prediction of proposed method, results of \citet{alfarraj2019semi} and \citet{blundell2015weight} are plotted in Figures \ref{pre-Marf}(c), \ref{pre-Marf}(e) and \ref{pre-Marf}(g) respectively. Meanwhile, the absolute difference between true AI and the results of the three methods are plotted in Figures \ref{pre-Marf}(d), \ref{pre-Marf}(f) and \ref{pre-Marf}(h) respectively. It can be seen that the modified network improves the prediction accuracy, especially for the gas channel around trace number 500. Moreover, our method reduces the artifacts in the AI image, which is the main advantage of using 2-$d$ CNN in pre-training.

We choose five different metrics to measure the performance of the three methods: mean squared error (MSE), Pearson’s correlation coefficient (PCC), coefficient of determination ($r^2$),  peak signal-to-noise ratio (PSNR/dB) and structural similarity index (SSIM) between true AI and the mean posterior prediction. The result is shown in Table \ref{pre-Mart}. All the metrics of proposed method show superiorities over another two methods.

\begin{table}[ht]
	\centering
  \caption{Performance metrics of different methods on Marmousi2 model.}\label{pre-Mart}
	\begin{tabular}[t]{cccccc}
		\toprule
		\diagbox[width=3cm]{\textbf{Methods}}{\textbf{Metric}} & \textbf{MSE} & ${\mathbf{PCC}}$ & \bm{$r^2$} & $\mathbf{PSNR}$ & $\mathbf{SSIM}$\\
		\midrule
		Proposed Method & 0.0387 & 0.9851 & 0.9556 & 28.9894 & 0.8896\\
		\citet{alfarraj2019semi} & 0.0581 & 0.9785 & 0.9333 & 27.2635 & 0.8483\\ 
		\citet{blundell2015weight} & 0.1577 & 0.9495 & 0.8232 & 22.8809 & 0.7558\\
		\bottomrule
	\end{tabular}
\end{table}

After pre-training, we estimate the uncertainty of the predictions on the Marmousi2 model. The prior variance $\sigma_0$ in \eqref{prior} is set to be $1\times 10^{-6}$, and the variance of the likelihood functions $\beta$ in \eqref{likelihood} is set to be 1. From the form of objective function \eqref{objective}, $\beta$ controls the trade-off between the prediction accuracy and the perturbation degree of network parameters. The number of sumpling $M$ in \eqref{expectation} is just set to be 1, in order to simplify the computation \cite{kingma2013auto}. In the prediction step, the number of sampling $N$ in \eqref{predictionMC} is set to be 40. The Bayesian neural network is trained for 3000 epochs.

\begin{figure}[!p]
  \setlength{\abovecaptionskip}{1cm}
	\begin{center}
    \subfigure[]{
      \includegraphics[width=0.7\textwidth,trim=80 100 180 130,clip]{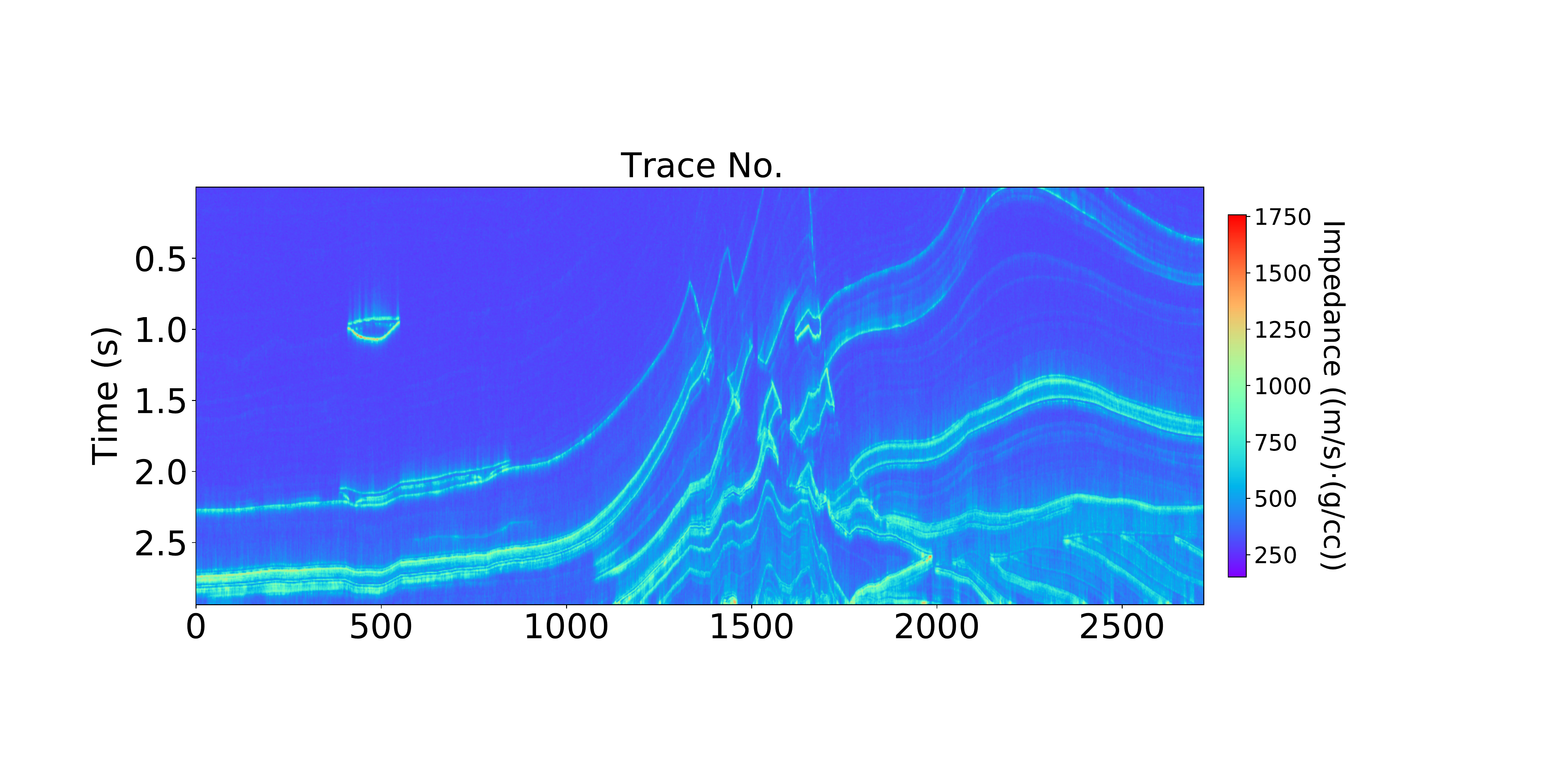} 
    } 
    \subfigure[]{
      \includegraphics[width=0.7\textwidth,trim=80 100 180 130,clip]{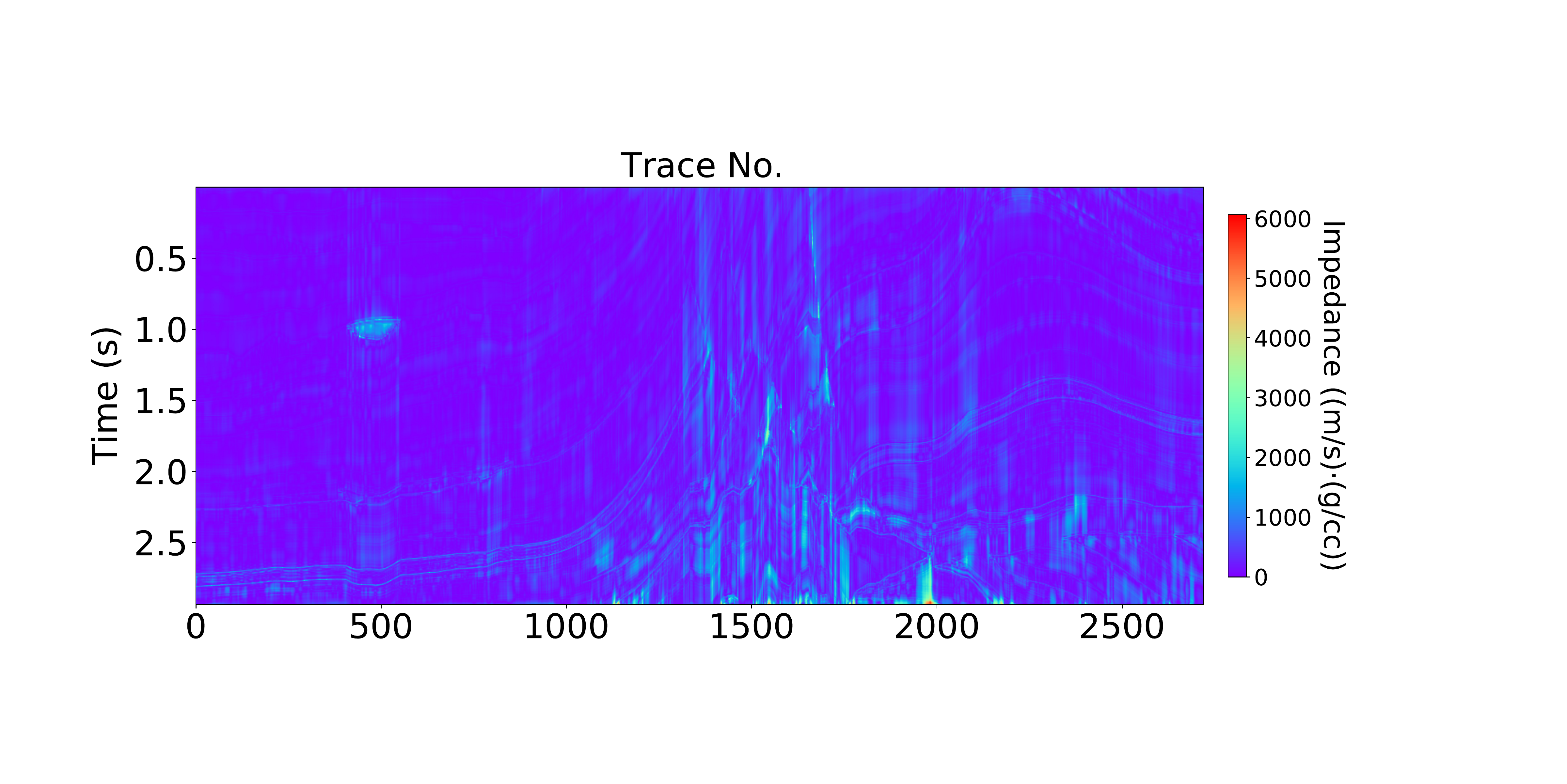}
    }
	  \caption{Comparision of prediction uncertainty and absolute difference of the Marmousi2 model test: (\textbf{a}) the prediction uncertainty; (\textbf{b}) impedance absolute difference of proposed method.}\label{un-Mar}
  \end{center}
\end{figure}

We use the standard variance of predictions to represent the uncertainty. The prediction uncertainty is shown in Figure \ref{un-Mar}(a), together with the absolute difference between true AI and the mean posterior prediction as is shown in Figure \ref{un-Mar}(b). From the figures, the prediction uncertainty indicates where the predictive deviations are relatively large. High uncertainty is observed mainly at impedance boundaries.

\begin{figure}[!p]
	\begin{center}
    \setlength{\abovecaptionskip}{1cm}
      \includegraphics[width=0.9\textwidth,trim=200 0 200 0,clip]{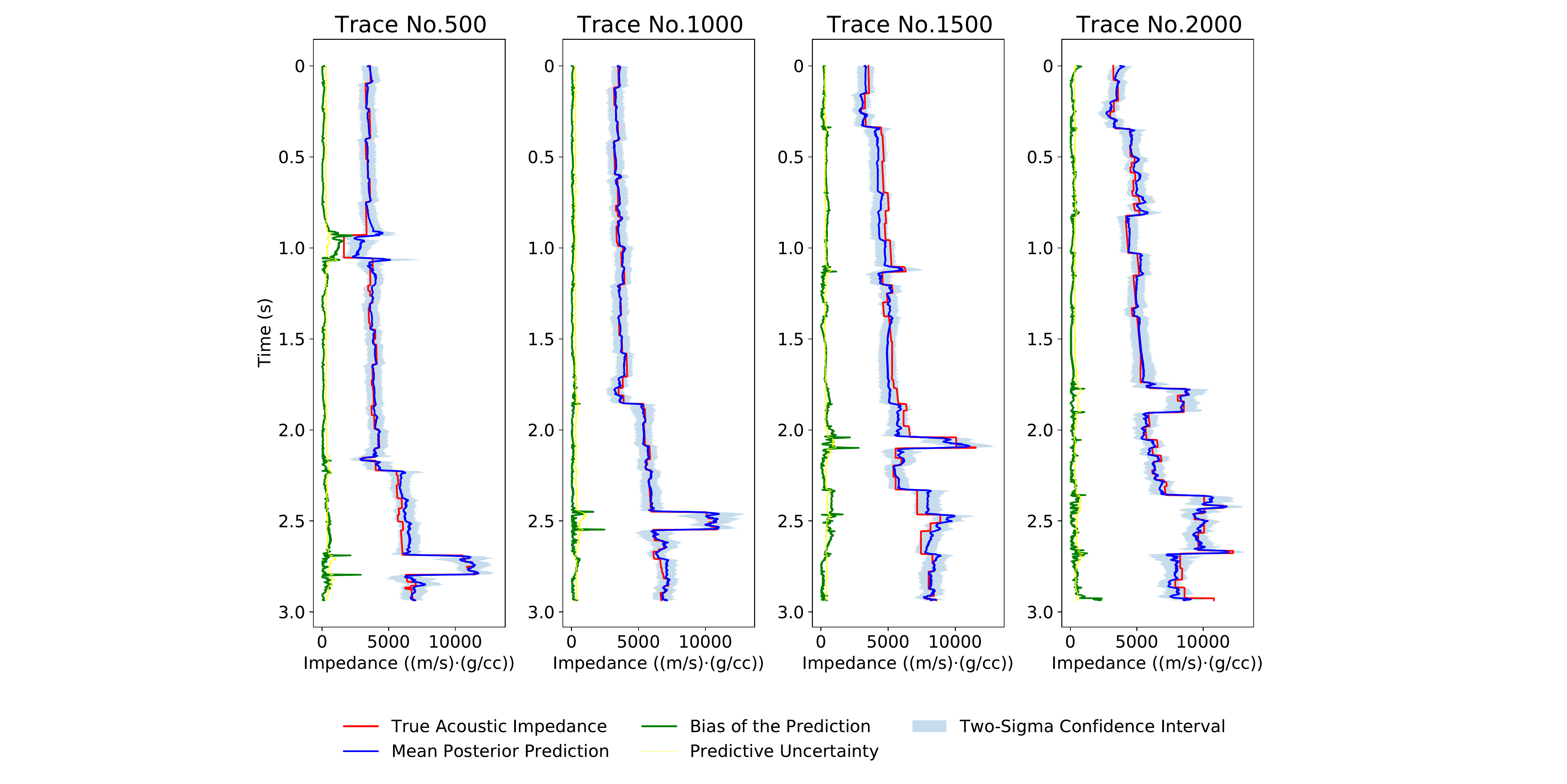}     
	  \caption{Prediction result of Marmousi2 model on Trace No. 500, 1000, 1500 and 2000.}\label{local-Mar}
  \end{center}
\end{figure}

The prediction result of Trace No. 500, 1000, 1500 and 2000 is shown in Figure \ref{local-Mar}, which shows the comparison of true AI, mean posterior prediction, prediction uncertainty and absolute difference. It also shows the two-sigma confidence interval $(x-2\sigma, x+2\sigma)$, where $\sigma$ is the standard variance of prediction. In Figure \ref{local-Mar}, most part of true AI is covered by $(x-2\sigma, x+2\sigma)$. The overall coverage of true AI is $96.33\%$, which is $93.29\%$ if using \citet{alfarraj2019semi} as pre-training, and is $10.75\%$ if using methods in \citet{blundell2015weight}. 

\subsection{SEAM model}
We further test the proposed strategy on SEAM model, which is widely used in geophysical studies \cite{mustafa2020spatiotemporal, cai2020wasserstein}. The data is from the open source code of \citet{mustafa2020spatiotemporal}, which contains seismic data of 501 traces, each has 701 time samples. The target AI has the same shape as the seismic data.

In pre-training, we choose 10 AI traces as target labels, also evenly distributed on the data field. The loss function are the same as \eqref{loss-Mar} in section \ref{Marmousi2 model}, except $N_l=10$ and $N_u=501$. Since the seismic data and AI have the same shape, the upsampling submodule in the inverse model is dropped, and the forward model is modified by decreasing the kernel size of convolutions. The inverse model and forward model are together trained for 1000 epochs.

\begin{figure}[!p]
  \setlength{\abovecaptionskip}{1cm}
	\begin{center}
    \subfigure[]{
      \includegraphics[width=0.48\textwidth,trim=200 40 180 50,clip]{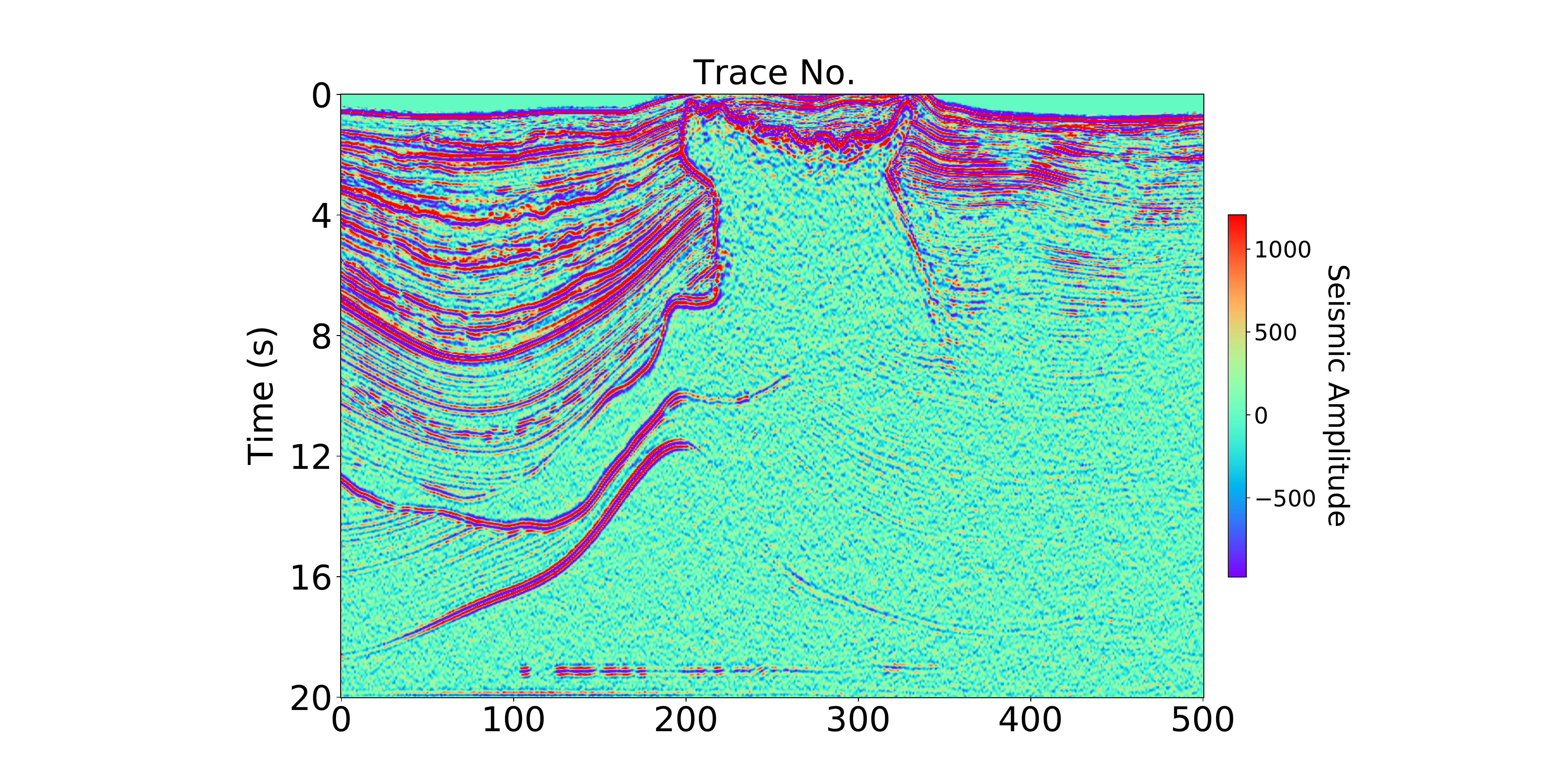} 
    }
    \subfigure[]{
      \includegraphics[width=0.48\textwidth,trim=200 40 180 50,clip]{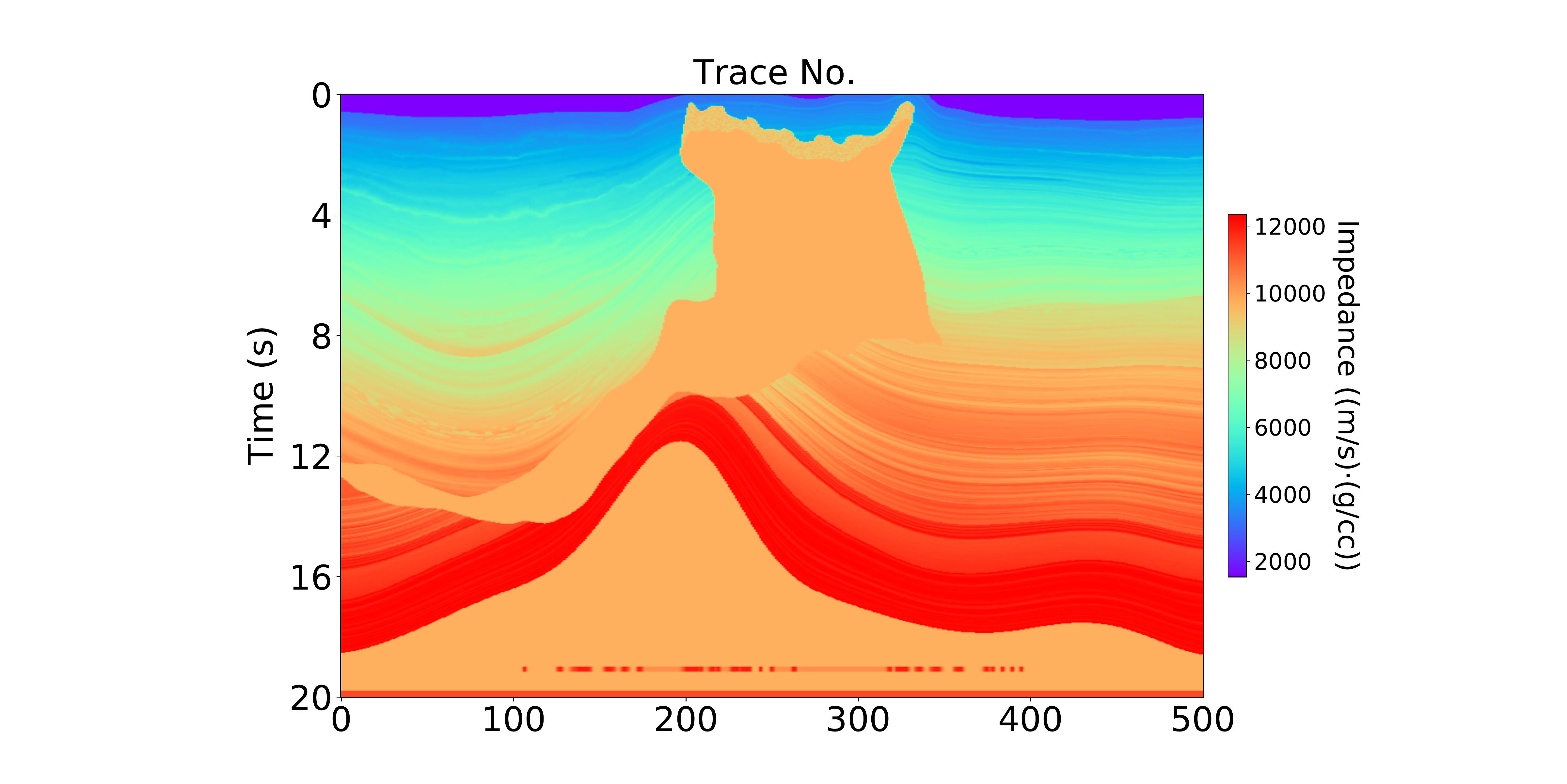}
    }
    \subfigure[]{
      \includegraphics[width=0.48\textwidth,trim=200 40 180 50,clip]{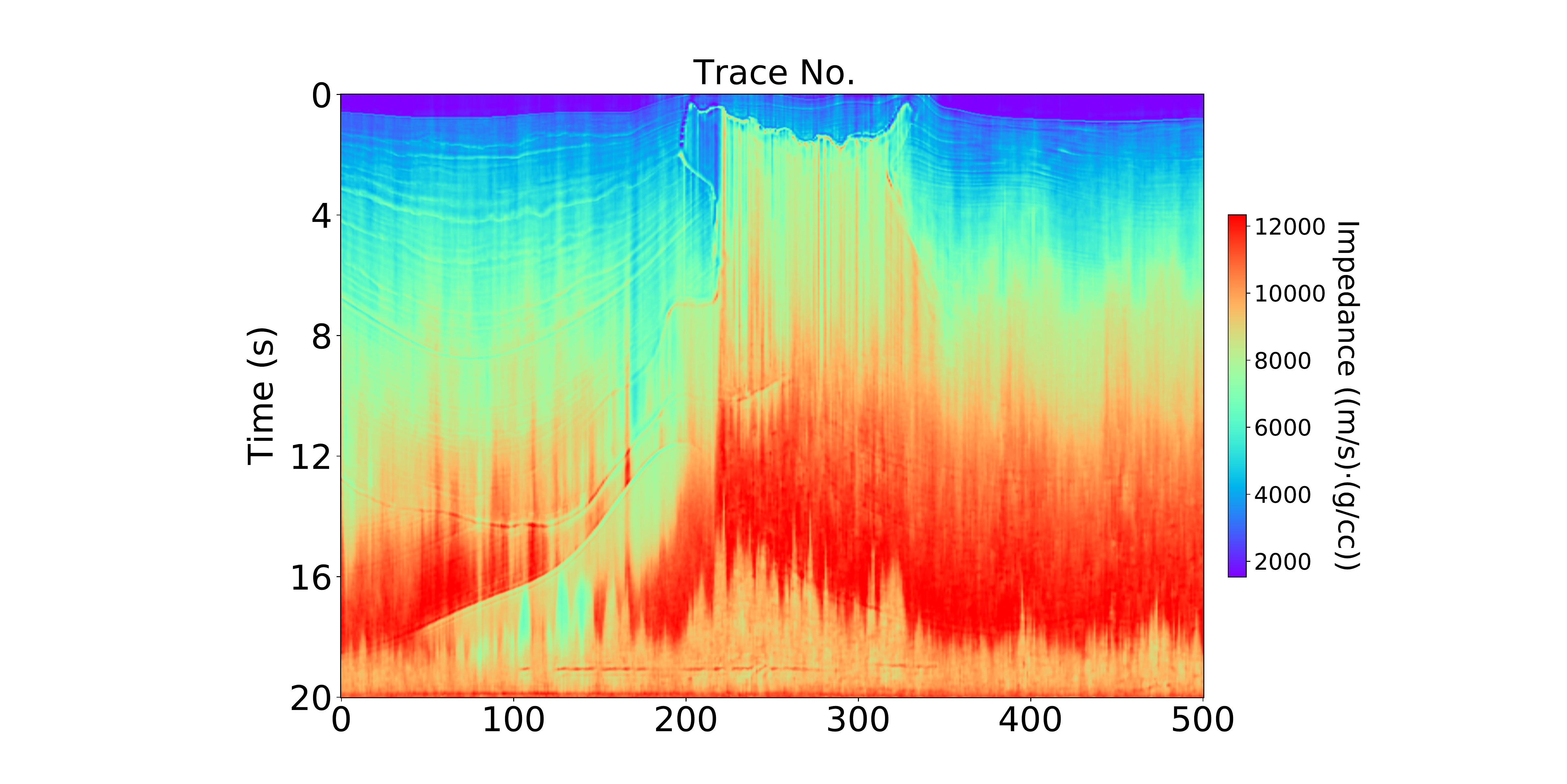}
    }
    \subfigure[]{
      \includegraphics[width=0.48\textwidth,trim=200 40 180 50,clip]{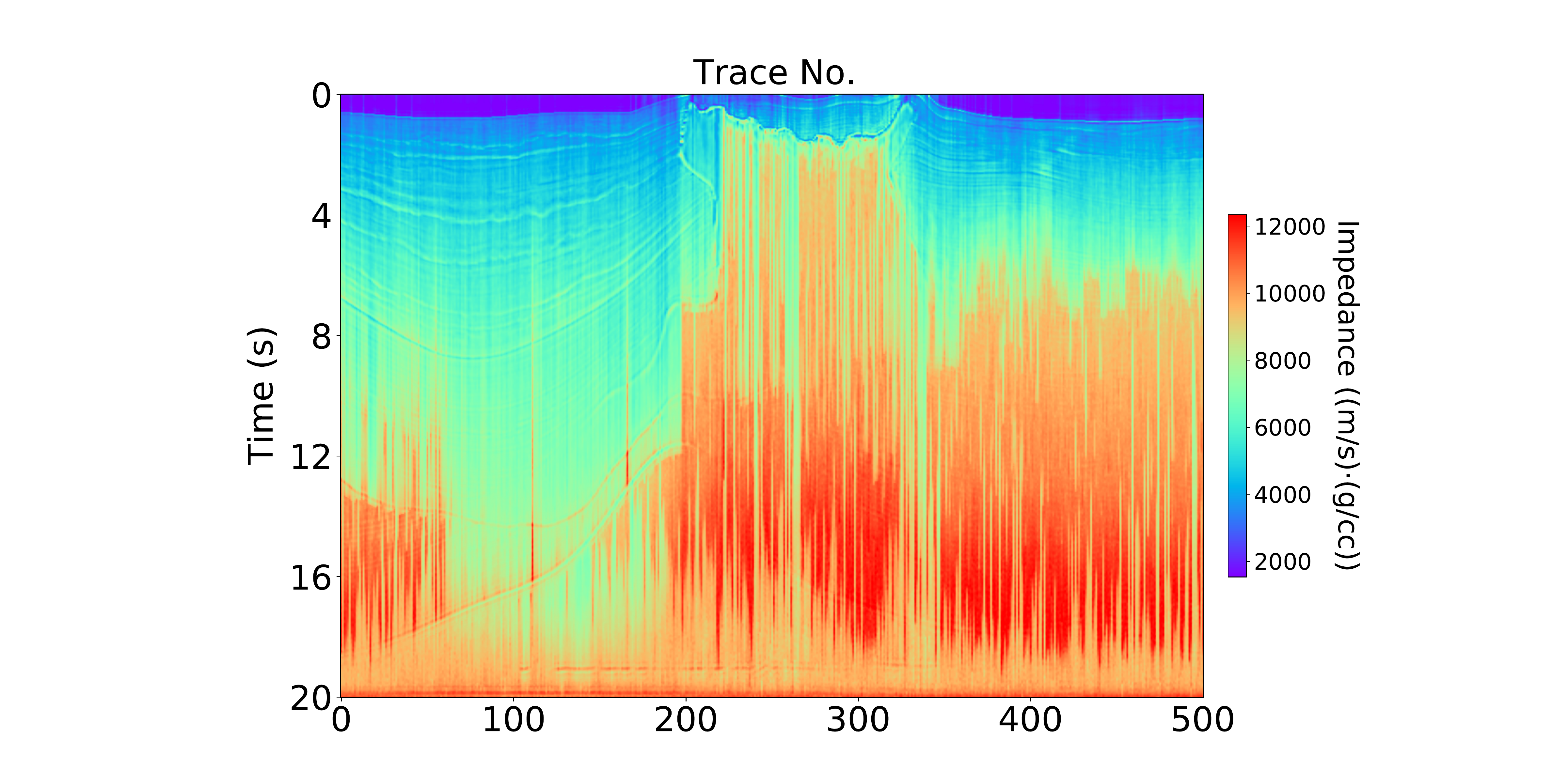}
    }
	  \caption{Seismic data, true AI, predicted AI and absolute difference on STAM model: (\textbf{a}) seismic data for inversion; (\textbf{b}) true AI; (\textbf{c}) AI predicted by proposed method; (\textbf{d}) AI predicted by \citet{alfarraj2019semi}.} \label{pre-SEAMf}
  \end{center}
\end{figure}

This time we reproduce the result of \citet*{alfarraj2019semi} as a baseline. From Figure \ref{pre-SEAMf}, the result predicted by our method is much more coherent horizontally. As in Table \ref{pre-SEAMt}, all the metrics of our method exceeds the baseline method on SEAM model.

\begin{table}[ht]
	\centering
  \caption{Performance metrics of different methods on SEAM model.}\label{pre-SEAMt}
	\begin{tabular}[t]{cccccc}
		\toprule
		\diagbox[width=3cm]{\textbf{Methods}}{\textbf{Metric}} & \textbf{MSE} & ${\mathbf{PCC}}$ & \bm{$r^2$} & $\mathbf{PSNR}$ & $\mathbf{SSIM}$\\
		\midrule
		Proposed Method & 0.1617 & 0.9187 & 0.7009 & 19.8696 & 0.5890\\
		\citet*{alfarraj2019semi} & 0.2809 & 0.8944 & 0.5850 & 17.4664 & 0.4165\\ 
		\bottomrule
	\end{tabular}
\end{table}

In uncertainty estimation, the prior variance $\sigma_0=1\times 10^{-8}$ and the variance of likelihood $\beta=1$.  The number of sumpling $M$ in \eqref{expectation} and $N$ in \eqref{predictionMC} is set to be 1 and 40 as in section \ref{Marmousi2 model}. The Bayesian neural network is trained for 1000 epochs.

\begin{figure}[!p]
  \setlength{\abovecaptionskip}{1cm}
	\begin{center}
    \subfigure[]{
      \includegraphics[width=0.6\textwidth,trim=200 40 180 50,clip]{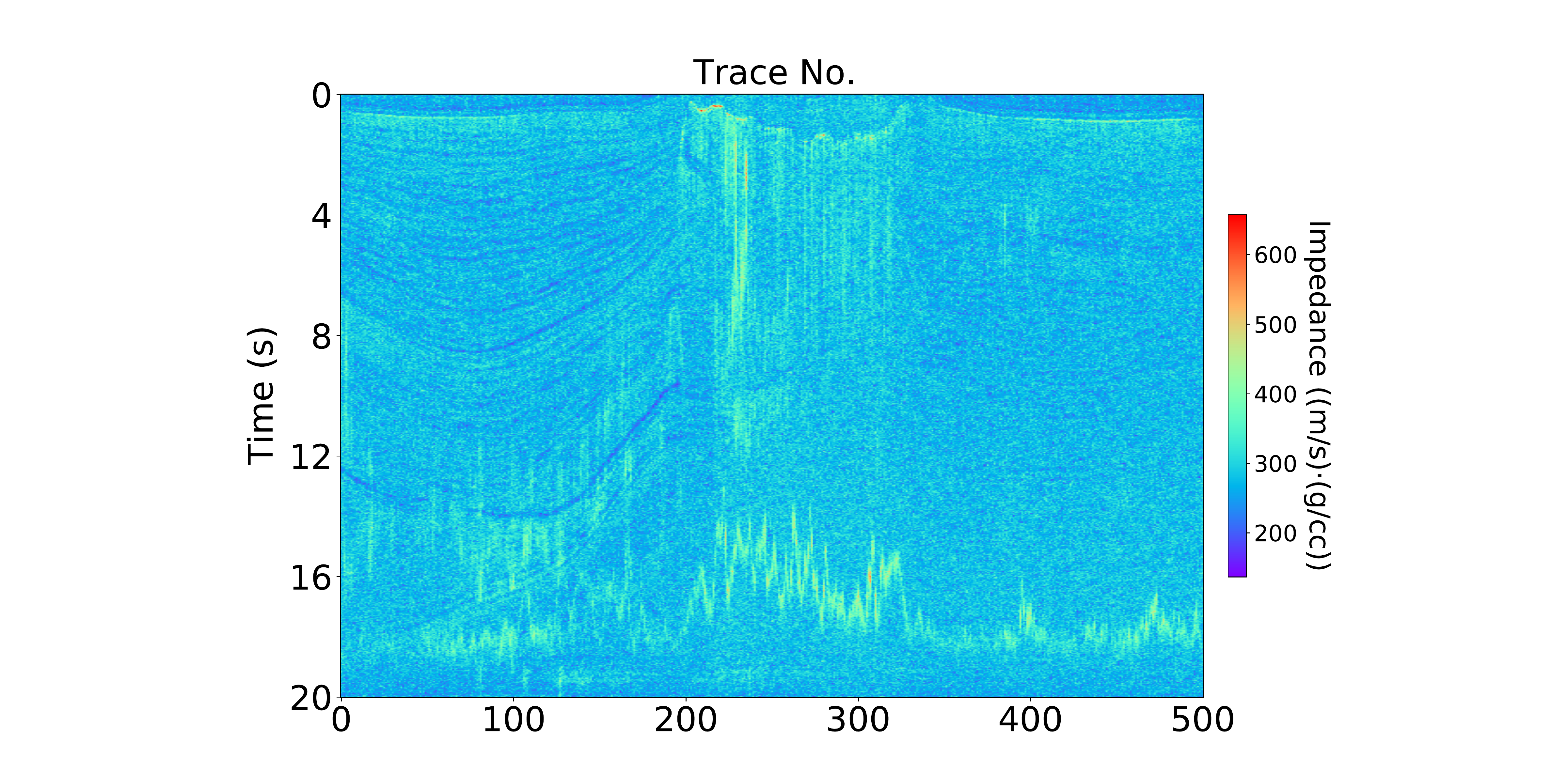} 
    } 
    \subfigure[]{
      \includegraphics[width=0.6\textwidth,trim=200 40 180 50,clip]{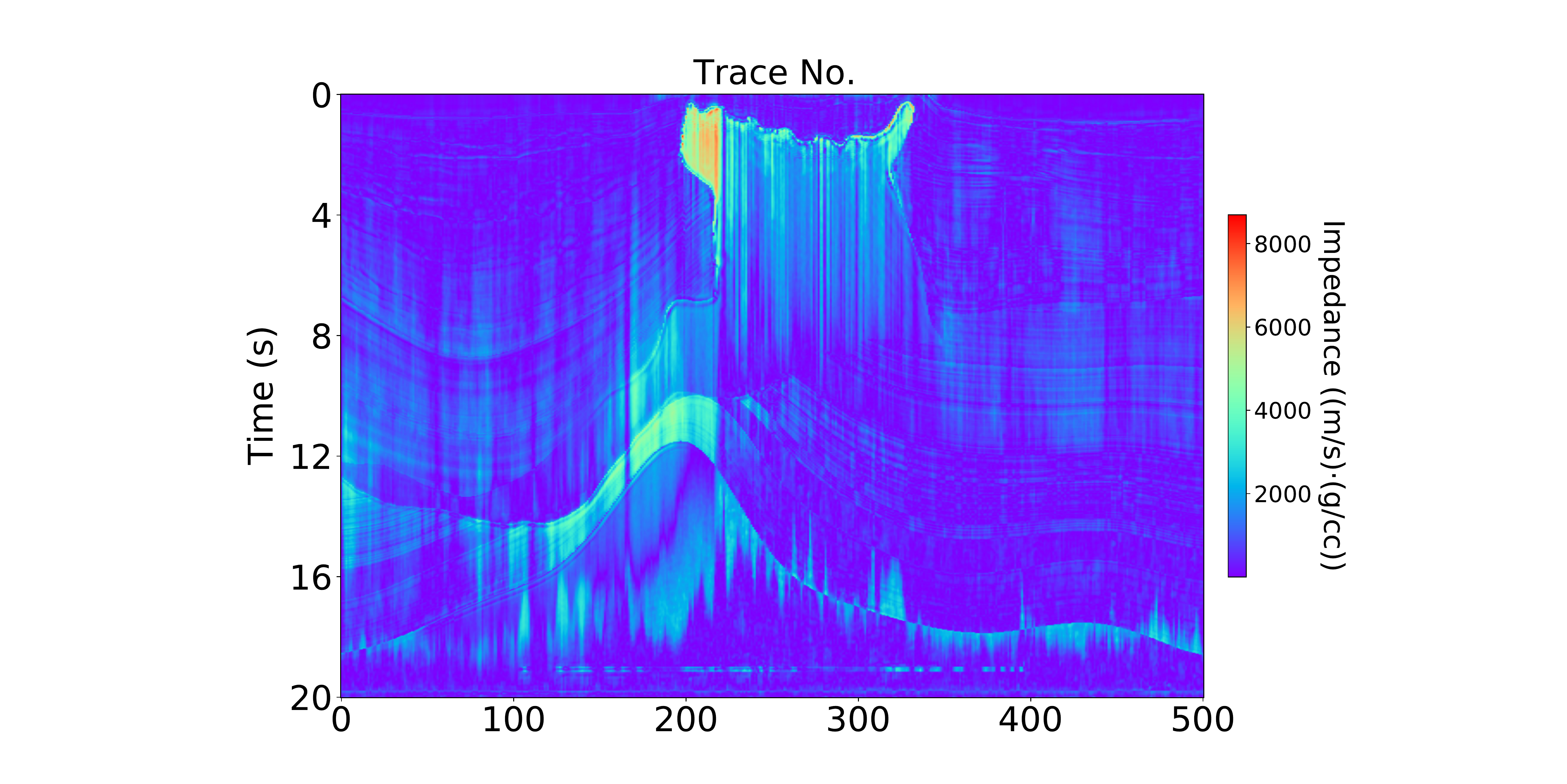}
    }
	  \caption{Comparision of prediction uncertainty and the absolute difference of the test on SEAM model: (\textbf{a}) the prediction uncertainty; (\textbf{b}) impedance absolute difference of proposed method.}\label{un-SEAM}
  \end{center}
\end{figure}

Figure \ref{un-SEAM}(a) shows the prediction uncertainty of proposed method, and Figure \ref{un-SEAM}(b) shows the absolute difference between true AI and mean posterior prediction. As in section \ref{Marmousi2 model}, the prediction uncertainty tends to increase where the mean posterior prediction diverges from true AI.

\subsection{Code availability}
An open-source implementation of our method can be downloaded from \url{https://github.com/Tom-900/Bayesian-Semi-supervised-Impedance-Inversion}.

\section{Conclusion}\label{Conclusion}

In this work, we use 2-$d$ CNN layers and 2-$d$ maxpooling layers in semi-supervised learning to improve the accuracy of impedance inversion result. The semi-supervised neural network is embedded into a Bayesian inference framework to estimate the prediction uncertainty. The hyperparameters in this framework are updated to minimize the KL divergence between approximated posterior distribution and the true posterior. Local reparameterization trick and linear approximation for GRU layers are used to reduce the computational cost. Numerical analysis based on Marmousi2 model and SEAM model demonstrates that our mean posterior prediction performs better in different measurements compared with original semi-supervised learning method. Furthermore, the prediction uncertainty indicates where the mean posterior prediction is relatively inaccurate. The way we used for uncertainty estimation has a concrete mathematical background and good interpretability from a Bayesian perspective.

\section{Acknowledgement}
 The authors would like to thank Ziyu Zhuang for useful discussion and comments. The authors declare that there is no conflict of interests regarding the publication of this article.

\bibliographystyle{plainnat}
\bibliography{reference.bib}

\begin{thebibliography}{34}
\providecommand{\natexlab}[1]{#1}
\providecommand{\url}[1]{\texttt{#1}}
\expandafter\ifx\csname urlstyle\endcsname\relax
  \providecommand{\doi}[1]{doi: #1}\else
  \providecommand{\doi}{doi: \begingroup \urlstyle{rm}\Url}\fi

\bibitem[Alfarraj and AlRegib(2018)]{alfarraj2018petrophysical}
Motaz Alfarraj and Ghassan AlRegib.
\newblock Petrophysical property estimation from seismic data using recurrent
  neural networks.
\newblock In \emph{2018 SEG International Exposition and Annual Meeting}.
  OnePetro, 2018.

\bibitem[Alfarraj and AlRegib(2019{\natexlab{a}})]{alfarraj2019semi}
Motaz Alfarraj and Ghassan AlRegib.
\newblock Semi-supervised learning for acoustic impedance inversion.
\newblock In \emph{SEG Technical Program Expanded Abstracts 2019}, pages
  2298--2302. Society of Exploration Geophysicists, 2019{\natexlab{a}}.

\bibitem[Alfarraj and AlRegib(2019{\natexlab{b}})]{alfarraj2019semisupervised}
Motaz Alfarraj and Ghassan AlRegib.
\newblock Semi-supervised sequence modeling for elastic impedance inversion.
\newblock \emph{Interpretation}, 7\penalty0 (3):\penalty0 SE237--SE249,
  2019{\natexlab{b}}.

\bibitem[Amini et~al.(2019)Amini, Schwarting, Soleimany, and
  Rus]{amini2019deep}
Alexander Amini, Wilko Schwarting, Ava Soleimany, and Daniela Rus.
\newblock Deep evidential regression.
\newblock \emph{arXiv preprint arXiv:1910.02600}, 2019.

\bibitem[Blundell et~al.(2015)Blundell, Cornebise, Kavukcuoglu, and
  Wierstra]{blundell2015weight}
Charles Blundell, Julien Cornebise, Koray Kavukcuoglu, and Daan Wierstra.
\newblock Weight uncertainty in neural network.
\newblock In \emph{International Conference on Machine Learning}, pages
  1613--1622. PMLR, 2015.

\bibitem[Buland and Omre(2003)]{buland2003bayesian}
Arild Buland and Henning Omre.
\newblock Bayesian linearized avo inversion.
\newblock \emph{Geophysics}, 68\penalty0 (1):\penalty0 185--198, 2003.

\bibitem[Cai et~al.(2020)Cai, Di, Li, Maniar, and Abubakar]{cai2020wasserstein}
Ao~Cai, Haibin Di, Zhun Li, Hiren Maniar, and Aria Abubakar.
\newblock Wasserstein cycle-consistent generative adversarial network for
  improved seismic impedance inversion: Example on 3d seam model.
\newblock In \emph{SEG Technical Program Expanded Abstracts 2020}, pages
  1274--1278. Society of Exploration Geophysicists, 2020.

\bibitem[Cho et~al.(2018)Cho, Gibson~Jr, and Zhu]{cho2018quasi}
Yongchae Cho, Richard~L Gibson~Jr, and Dehan Zhu.
\newblock Quasi 3d transdimensional markov-chain monte carlo for seismic
  impedance inversion and uncertainty analysis.
\newblock \emph{Interpretation}, 6\penalty0 (3):\penalty0 T613--T624, 2018.

\bibitem[Choi et~al.(2020)Choi, Kim, and Byun]{choi2020uncertainty}
Junhwan Choi, Dowan Kim, and Joongmoo Byun.
\newblock Uncertainty estimation in impedance inversion using bayesian deep
  learning.
\newblock In \emph{SEG International Exposition and Annual Meeting}. OnePetro,
  2020.

\bibitem[Das et~al.(2019)Das, Pollack, Wollner, and
  Mukerji]{das2019convolutional}
Vishal Das, Ahinoam Pollack, Uri Wollner, and Tapan Mukerji.
\newblock Convolutional neural network for seismic impedance inversion.
\newblock \emph{Geophysics}, 84\penalty0 (6):\penalty0 R869--R880, 2019.

\bibitem[Di and Abubakar(2021)]{di2021estimating}
Haibin Di and Aria Abubakar.
\newblock Estimating subsurface properties using a semi-supervised neural
  networks approach.
\newblock \emph{Geophysics}, 87\penalty0 (1):\penalty0 1--38, 2021.

\bibitem[Di et~al.(2020)Di, Chen, Maniar, and Abubakar]{di2020semi}
Haibin Di, Xiaoli Chen, Hiren Maniar, and Aria Abubakar.
\newblock Semi-supervised seismic and well log integration for reservoir
  property estimation.
\newblock In \emph{SEG International Exposition and Annual Meeting}. OnePetro,
  2020.

\bibitem[Esposito(2020)]{esposito2020blitzbdl}
Piero Esposito.
\newblock Blitz - bayesian layers in torch zoo (a bayesian deep learing library
  for torch).
\newblock \url{https://github.com/piEsposito/blitz-bayesian-deep-learning/},
  2020.

\bibitem[Gal and Ghahramani(2015)]{gal2015bayesian}
Yarin Gal and Zoubin Ghahramani.
\newblock Bayesian convolutional neural networks with bernoulli approximate
  variational inference.
\newblock \emph{arXiv preprint arXiv:1506.02158}, 2015.

\bibitem[Gal and Ghahramani(2016{\natexlab{a}})]{gal2016dropout}
Yarin Gal and Zoubin Ghahramani.
\newblock Dropout as a bayesian approximation: Representing model uncertainty
  in deep learning.
\newblock In \emph{international conference on machine learning}, pages
  1050--1059. PMLR, 2016{\natexlab{a}}.

\bibitem[Gal and Ghahramani(2016{\natexlab{b}})]{gal2016theoretically}
Yarin Gal and Zoubin Ghahramani.
\newblock A theoretically grounded application of dropout in recurrent neural
  networks.
\newblock \emph{Advances in neural information processing systems},
  29:\penalty0 1019--1027, 2016{\natexlab{b}}.

\bibitem[Iturrar{\'a}n-Viveros et~al.(2021)Iturrar{\'a}n-Viveros,
  Mu{\~n}oz-Garc{\'\i}a, Castillo-Reyes, and Shukla]{iturraran2021machine}
Ursula Iturrar{\'a}n-Viveros, Andr{\'e}s~M Mu{\~n}oz-Garc{\'\i}a, Octavio
  Castillo-Reyes, and Khemraj Shukla.
\newblock Machine learning as a seismic prior velocity model building method
  for full-waveform inversion: a case study from colombia.
\newblock \emph{Pure and Applied Geophysics}, 178\penalty0 (2):\penalty0
  423--448, 2021.

\bibitem[Kim and Nakata(2018)]{kim2018geophysical}
Yuji Kim and Nori Nakata.
\newblock Geophysical inversion versus machine learning in inverse problems.
\newblock \emph{The Leading Edge}, 37\penalty0 (12):\penalty0 894--901, 2018.

\bibitem[Kingma and Welling(2013)]{kingma2013auto}
Diederik~P Kingma and Max Welling.
\newblock Auto-encoding variational bayes.
\newblock \emph{arXiv preprint arXiv:1312.6114}, 2013.

\bibitem[Kingma et~al.(2015)Kingma, Salimans, and
  Welling]{kingma2015variational}
Durk~P Kingma, Tim Salimans, and Max Welling.
\newblock Variational dropout and the local reparameterization trick.
\newblock \emph{Advances in neural information processing systems},
  28:\penalty0 2575--2583, 2015.

\bibitem[Li and Peng(2017)]{li2017seismic}
Shu Li and Zhenming Peng.
\newblock Seismic acoustic impedance inversion with multi-parameter
  regularization.
\newblock \emph{Journal of Geophysics and Engineering}, 14\penalty0
  (3):\penalty0 520--532, 2017.

\bibitem[Ma et~al.(2021)Ma, Wang, Wang, and Lu]{ma2021improved}
Qiming Ma, Yuqing Wang, Qi~Wang, and Wenkai Lu.
\newblock Improved seismic impedance inversion based on uncertainty analysis.
\newblock In \emph{First International Meeting for Applied Geoscience \&
  Energy}, pages 1400--1404. Society of Exploration Geophysicists, 2021.

\bibitem[Martin et~al.(2002)Martin, Marfurt, and Larsen]{martin2002marmousi}
Gary~S Martin, Kurt~J Marfurt, and Shawn Larsen.
\newblock Marmousi-2: An updated model for the investigation of avo in
  structurally complex areas.
\newblock In \emph{SEG Technical Program Expanded Abstracts 2002}, pages
  1979--1982. Society of Exploration Geophysicists, 2002.

\bibitem[Meng et~al.(2020)Meng, Wu, Liu, and Chen]{meng2020semi}
Delin Meng, Bangyu Wu, Naihao Liu, and Wenchao Chen.
\newblock Semi-supervised deep learning seismic impedance inversion using
  generative adversarial networks.
\newblock In \emph{IGARSS 2020-2020 IEEE International Geoscience and Remote
  Sensing Symposium}, pages 1393--1396. IEEE, 2020.

\bibitem[Meng et~al.(2021)Meng, Wu, Wang, and Zhu]{meng2021seismic}
Delin Meng, Bangyu Wu, Zhiguo Wang, and Zhaolin Zhu.
\newblock Seismic impedance inversion using conditional generative adversarial
  network.
\newblock \emph{IEEE Geoscience and Remote Sensing Letters}, 2021.

\bibitem[Mustafa and AlRegib(2020)]{mustafa2020joint}
Ahmad Mustafa and Ghassan AlRegib.
\newblock Joint learning for seismic inversion: An acoustic impedance
  estimation case study.
\newblock In \emph{SEG Technical Program Expanded Abstracts 2020}, pages
  1686--1690. Society of Exploration Geophysicists, 2020.

\bibitem[Mustafa et~al.(2020)Mustafa, Alfarraj, and
  AlRegib]{mustafa2020spatiotemporal}
Ahmad Mustafa, Motaz Alfarraj, and Ghassan AlRegib.
\newblock Spatiotemporal modeling of seismic images for acoustic impedance
  estimation.
\newblock In \emph{SEG International Exposition and Annual Meeting}. OnePetro,
  2020.

\bibitem[Nunes et~al.(2019)Nunes, Azevedo, and Soares]{nunes2019fast}
Ruben Nunes, Leonardo Azevedo, and Am{\'\i}lcar Soares.
\newblock Fast geostatistical seismic inversion coupling machine learning and
  fourier decomposition.
\newblock \emph{Computational Geosciences}, 23\penalty0 (5):\penalty0
  1161--1172, 2019.

\bibitem[Russell(2019)]{russell2019machine}
Brian Russell.
\newblock Machine learning and geophysical inversion—a numerical study.
\newblock \emph{The Leading Edge}, 38\penalty0 (7):\penalty0 512--519, 2019.

\bibitem[Siahkoohi et~al.(2020)Siahkoohi, Rizzuti, and
  Herrmann]{siahkoohi2020deep}
Ali Siahkoohi, Gabrio Rizzuti, and F~Herrmann.
\newblock A deep-learning based bayesian approach to seismic imaging and
  uncertainty quantification.
\newblock In \emph{82nd EAGE Annual Conference \& Exhibition}, volume 2020,
  pages 1--5. European Association of Geoscientists \& Engineers, 2020.

\bibitem[Tarantola(1986)]{tarantola1986strategy}
Albert Tarantola.
\newblock A strategy for nonlinear elastic inversion of seismic reflection
  data.
\newblock \emph{Geophysics}, 51\penalty0 (10):\penalty0 1893--1903, 1986.

\bibitem[Wu et~al.(2020)Wu, Meng, Wang, Liu, and Wang]{wu2020seismic}
Bangyu Wu, Delin Meng, Lingling Wang, Naihao Liu, and Ying Wang.
\newblock Seismic impedance inversion using fully convolutional residual
  network and transfer learning.
\newblock \emph{IEEE Geoscience and Remote Sensing Letters}, 17\penalty0
  (12):\penalty0 2140--2144, 2020.

\bibitem[Wu et~al.(2021{\natexlab{a}})Wu, Meng, and Zhao]{wu2021semi}
Bangyu Wu, Delin Meng, and Haixia Zhao.
\newblock Semi-supervised learning for seismic impedance inversion using
  generative adversarial networks.
\newblock \emph{Remote Sensing}, 13\penalty0 (5):\penalty0 909,
  2021{\natexlab{a}}.

\bibitem[Wu et~al.(2021{\natexlab{b}})Wu, Yan, Bi, Zhang, and Si]{wu2021deep}
Xinming Wu, Shangsheng Yan, Zhengfa Bi, Sibo Zhang, and Hongjie Si.
\newblock Deep learning for multi-dimensional seismic impedance inversion.
\newblock \emph{Geophysics}, 86\penalty0 (5):\penalty0 1--44,
  2021{\natexlab{b}}.

\end{thebibliography}

\clearpage
\appendix
\clearpage

\clearpage

\clearpage

\clearpage

\clearpage

\clearpage

\clearpage

\clearpage

\end{document}